\documentclass[aps,pra,10pt,twocolumn,floatfix]{revtex4-2}

\usepackage[T1]{fontenc}
\usepackage[colorlinks=true,urlcolor=blue,linkcolor=blue,citecolor=blue]{hyperref}
\usepackage{graphicx,overpic,orcidlink,amssymb,bm,physics,siunitx,soul,dsfont}

\newcommand{\orcG}{\orcidlink{0000-0002-2335-574X}}
\newcommand{\orcC}{\orcidlink{0000-0002-9054-4225}}
\newcommand{\orcM}{\orcidlink{0000-0002-2636-9936}}
\newcommand{\emaG}{\email{guillermo.preisser@quantinuum.com}}
\newcommand{\emaM}{\email{michael.lubasch@quantinuum.com}}
\newcommand{\aff}{\affiliation{Quantinuum, Partnership House, Carlisle Place, London SW1P 1BX, United Kingdom}}

\begin{document}

\title{Variational matrix product states for combinatorial optimization}

\author{Guillermo Preisser\orcG}\emaG\aff
\author{Conor Mc Keever\orcC}\aff
\author{Michael Lubasch\orcM}\emaM\aff

\date{July 5, 2026}

\begin{abstract}
To compute approximate solutions for combinatorial optimization problems, we describe variational methods based on the product state (PS) and matrix product state (MPS) ans\"{a}tze.
We perform variational energy minimization with respect to a quantum annealing Hamiltonian and utilize randomness by embedding the approaches in the metaheuristic iterated local search (ILS).
The resulting quantum-inspired ILS algorithms are benchmarked on maximum cut problems of up to \num{50000} variables.
We show that they can outperform traditional (M)PS methods, classical ILS, the quantum approximate optimization algorithm and other variational quantum-inspired solvers.
\end{abstract}

\maketitle

\section{Introduction}

Remarkable algorithmic discoveries have enabled combinatorial optimization to solve problems of relevance to several industries, including manufacturing~\cite{ZhEtAl23}, transportation~\cite{GuHa08, FaEtAl13} and telecommunication~\cite{LiEtAl24}.
A key ingredient of certain successful classical approaches, such as iterated local search (ILS)~\cite{LoMaSt19}, is the strategic use of randomness to escape local optima and efficiently explore the solution space in search of better optima.
From the perspective of quantum computing, promising proposals for combinatorial optimization are quantum annealing~\cite{KaNi98, FaEtAl00, DaCh08, AlLi18, HaEtAl20} and variational quantum algorithms~\cite{CeEtAl21, BhEtAl22, TiEtAl22}, in particular the quantum approximate optimization algorithm (QAOA)~\cite{FaGoGu14, ZhEtAl20}.
However, current quantum hardware suffers from experimental noise to such an extent that certain quantum computations can be amenable to classical simulations, for which the variational matrix product state (MPS) ansatz~\cite{PeEtAl07, VeMuCi08, Sc11, Or14, Ba23} has proven to be a particularly powerful tool~\cite{ZhStWa20, AyEtAl23}.
This naturally raises the following question:
By making use of both classical and quantum concepts, can MPS algorithms outperform established classical solvers for combinatorial optimization?

In this paper, we combine MPS algorithms for quantum annealing with ILS and show that the resulting quantum-inspired ILS (QiILS) approaches outperform both classical ILS and standard MPS methods.
In the context of maximum cut (MaxCut) problems~\cite{Ka72, GoWi95}, which are a paradigmatic benchmark in combinatorial optimization, we also show that QiILS can find better solutions more efficiently than QAOA, the variational quantum-inspired algorithm local quantum annealing (LQA)~\cite{BoEtAl22} and its variant based on generalized group-theoretic coherent states (GCS)~\cite{GuEtAl21, ScEtAl22, FiSa25} (see Appendix~\ref{app:OverviewOfBenchmarkOptimizationMethods} for concise descriptions of ILS, LQA and GCS).
Furthermore, we introduce an embarrassingly parallel alternative to QiILS:
Quantum-inspired iterative global search (QiIGS) employs global gradient updates that enable parallelization across problem size.
We show that QiIGS can efficiently utilize GPU-based parallel computing, exhibit a runtime nearly independent of problem size for instances with up to \num{50000} variables and provide an order-of-magnitude speedup over QiILS for the largest, \num{20000}-variable, MaxCut instance (G81) from the Gset dataset~\cite{Gset}.

\begin{figure}
\centering
\includegraphics[width=\columnwidth]{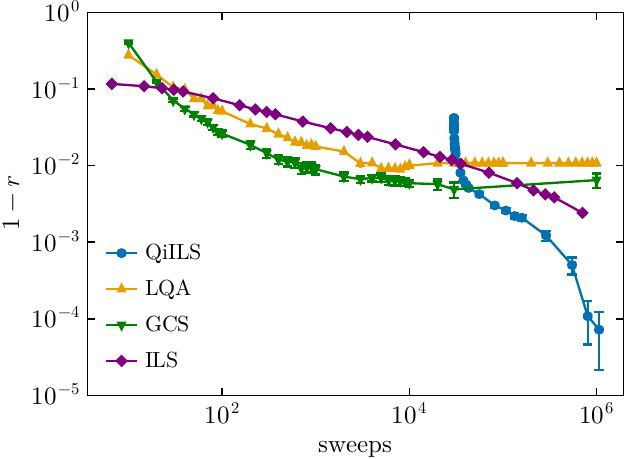}
\caption{\label{fig:1}
Performance comparison for G12.
Average relative error $1 - r$, averaged over 100 random initial states, for G12~\cite{Gset}, a toroidal weighted MaxCut problem of 800 variables and weights from $\{-1, 1\}$.
If not visible, error bars are smaller than their associated symbols.
The offset for QiILS results from additional sweeps used to tune a hyperparameter that aims to obtain more consistent and faster improvement per sweep.
Despite this offset, QiILS achieves the best performance.
For each method, results correspond to the best hyperparameter choice within the explored hyperparameter search space.
Each sweep of QiILS and ILS runs in a similar amount of time, whereas LQA (GCS) is roughly 7 (\num{9000}) times slower.
Further details on hyperparameters as well as runtimes per sweep are provided in Appendixes~\ref{app:HyperparametersUsed} and~\ref{app:TimesPerSweepForEachMethod}.
}
\end{figure}

Figure~\ref{fig:1} illustrates the typical advantage of QiILS over other solvers on the Gset problems.
In this example, we compare QiILS with LQA, GCS, and ILS on G12, a standard benchmark graph (see figure caption for details).
Performance is evaluated in terms of the relative error, defined as $1-r$, where $r$ denotes the approximation ratio between the obtained and optimal (or best known) cut values.
After an initialization phase of $\approx 10^{4}$ sweeps, used to tune a hyperparameter, QiILS consistently outperforms the competing methods and achieves more than an order-of-magnitude improvement.

Our study is motivated by earlier works on discrete optimization using (M)PSs~\cite{BaEtAl06, SmSm14, BaEtAl15, HaMo18, MuEtAl22, SrEtAl22, DuEtAl22a, VeVa22, DuEtAl22b, BoEtAl22, LaEtAl23, LoChPe23, BaEtAl23, AlEtAl24, GaLo24, LoCh25, NaTaTa25, MoEtAl25, RaEtAl26} and more complex tensor networks~\cite{GaLa12, BaEtAl15, BiMoTu15, ZhKa19, KoEtAl19, LiWaZh21, RaEtAl21, HaEtAl22, LiEtAl23, PaGr23, YaSoMi24, GaGr24, PaSiOr25, Ma25, DzEtAl25, TeEtAl26}.
Most closely related to our proposals are approaches that propagate (M)PSs along adiabatic paths~\cite{BaEtAl06, SmSm14, BaEtAl15, HaMo18, VeVa22, BoEtAl22, LaEtAl23, RaEtAl26} or in imaginary time~\cite{MuEtAl22}.
Compared with these prior methods, our algorithms more directly target the solution space and deliberately exploit randomness to find better solutions.
They also facilitate a comparison of the utility of randomness and limited quantum entanglement (accessible via MPS) in combinatorial optimization.

The paper is organized as follows.
Section~\ref{sec:Background} contains background information.
We present and analyze QiILS based on MPS and product states (PSs) in Secs.~\ref{sec:QiILS} and~\ref{sec:QiILSForPSs}, respectively.
Then, we investigate the performance mechanism underlying QiILS in Sec.~\ref{sec:PerformanceMechanism}, discuss the selection of the main QiILS hyperparameter in Sec.~\ref{sec:SelectionOfHyperparameterLambda}, perform a comparison with QAOA in Sec.~\ref{sec:ComparisonWithQAOA}, present additional Gset results in Sec.~\ref{sec:GsetBenchmark}, and introduce and study QiIGS in Sec.~\ref{sec:QiIGS}.
Finally, we conclude with Sec.~\ref{sec:Conclusions}.
Additional information is provided in the Appendixes.

\section{Background}
\label{sec:Background}

For a quantum system of $n$ qubits, an MPS with open boundary conditions takes the form $\ket{\Psi} = \sum_{\{s_j\}} A^{s_1}_1 A^{s_2}_2 \ldots A^{s_n}_n \ket{s_1, s_2, \ldots, s_n}$~\cite{PeEtAl07} where $s_j \in \{0, 1\}$, each $A^{s_j}_j$ is a matrix of size $\chi_{j-1} \times \chi_j$, $\chi_0 = 1 = \chi_n$, and $\chi_j = \chi$ for $0 < j < n$.
The so-called bond dimension $\chi$ determines the maximum amount of entanglement the MPS can capture.
For $\chi = 1$, the MPS represents an unentangled PS.
Larger $\chi$ increases the expressivity of the MPS ansatz but also leads to higher computational costs.
Throughout this work, we compute approximate ground states using the variational MPS implementation of the density matrix renormalization group (DMRG) algorithm~\cite{VeMuCi08, Sc11, Or14, Ba23} provided by the ITensor library~\cite{FiWhSt22a, FiWhSt22b}.
For the quantum annealing Hamiltonians considered in this paper, we use matrix product operator (MPO) representations~\cite{VeMuCi08, Sc11, Or14, Ba23} within DMRG that are numerically calculated by ITensor routines; however, it is worth highlighting that optimal analytical MPO constructions for these types of Hamiltonians exist~\cite{FrNeDu10}.

The MaxCut problem~\cite{Ka72, GoWi95} is defined as follows.
We assume that a graph is given with vertices $\{1, 2, \ldots, n\}$, edges $\{(j, k): w_{j, k} \neq 0\}$, and weights $w_{j, k}$ between vertices $j$ and $k$.
The goal is to find a partition of the vertices into two disjoint subsets (equivalently, an assignment of spin values $\in \{-1, 1\}$ to the vertices) that maximizes the weight of the cut, i.e.\ the sum of the weights associated with the edges crossing the partition.
We consider three types of graphs: unweighted $d$-regular graphs (udR) where $w_{j, k} = 1$, weighted $d$-regular graphs (wdR) where $w_{j, k}$ are drawn uniformly at random from $[0, 1]$, and Gset instances~\cite{Gset} where $w_{j, k} \in \{-1, 1\}$.

To evaluate performance, we use the so-called approximation ratio, defined as
\begin{equation}
 r = \frac{C(\mathbf{b})}{C_{\max}},
\end{equation}
where $\mathbf{b} = (b_1, b_2, \ldots, b_n)$ is a bitstring of bit values $b_j \in \{0, 1\}$ representing the spin values of the vertices, $C(\mathbf{b})$ is the total weight of edges crossing the partition corresponding to bitstring $\mathbf{b}$, and $C_{\max} = \max_{\mathbf{b} \in \{0, 1\}^n} C(\mathbf{b})$ denotes the optimal cut value.

A common approach to solving combinatorial optimization problems with quantum methods maps the classical problem onto a quantum spin system.
In this mapping, each classical bit $b_j$ corresponds to the eigenvalues of the Pauli-$Z$ operator $\sigma^Z$ and the optimization objective is encoded in an Ising Hamiltonian $H_\text{f} = \sum_{\langle j, k \rangle} w_{j, k} \, \sigma_j^Z \sigma_k^Z$.
Here, the couplings $w_{j, k}$ define the edges of the underlying graph, making the correspondence between spin interactions and the original combinatorial structure explicit.
The cut value $C$ is directly related to the ground state energy $E$ of $H_\text{f}$ via $C = (\sum_{j<k} w_{j, k} - E) / 2$~\cite{BaEtAl88, Lu14}.

\section{Quantum-inspired iterated local search (QiILS)}
\label{sec:QiILS}

To approximate the ground state of the Ising Hamiltonian within the MPS framework, we employ techniques inspired by ILS~\cite{LoMaSt19} and quantum annealing~\cite{KaNi98, FaEtAl00, DaCh08, AlLi18, HaEtAl20}.
In particular, we consider an interpolation between the initial Hamiltonian $H_\text{i} = -\sum_{j=1}^n \sigma_j^X$ with Pauli-$X$ operator $\sigma^X$ and the problem Hamiltonian:
\begin{equation}\label{eq:adiab}
 H(\lambda) = (1-\lambda) H_\text{i} + \lambda H_\text{f}, \quad \lambda \in [0,1].
\end{equation}
Using the Hamiltonian of Eq.~\eqref{eq:adiab} at a fixed value of $\lambda$, we perform an iterative procedure indexed by $\iota$, where each iteration consists of a DMRG optimization followed by a perturbation step, in analogy with ILS.
Starting from a randomly initialized MPS with bond dimension $\chi$, we perform DMRG sweeps at the chosen $\lambda$ until convergence.
Once converged, we generate a spin configuration (bitstring solution) $\mathbf{b}$ according to the probability distribution encoded in the MPS.
The configuration is constructed sequentially by selecting, at each site, the spin value most probable given the previously chosen spins --- a greedy procedure similar to that used in Refs.~\cite{SrEtAl22, ChEtAl22}.
We evaluate the corresponding energy with respect to the spin Hamiltonian $H_\text{f}$ and keep the bitstring if it yields the lowest energy found so far.

\begin{figure}
\centering
\includegraphics[width=\columnwidth]{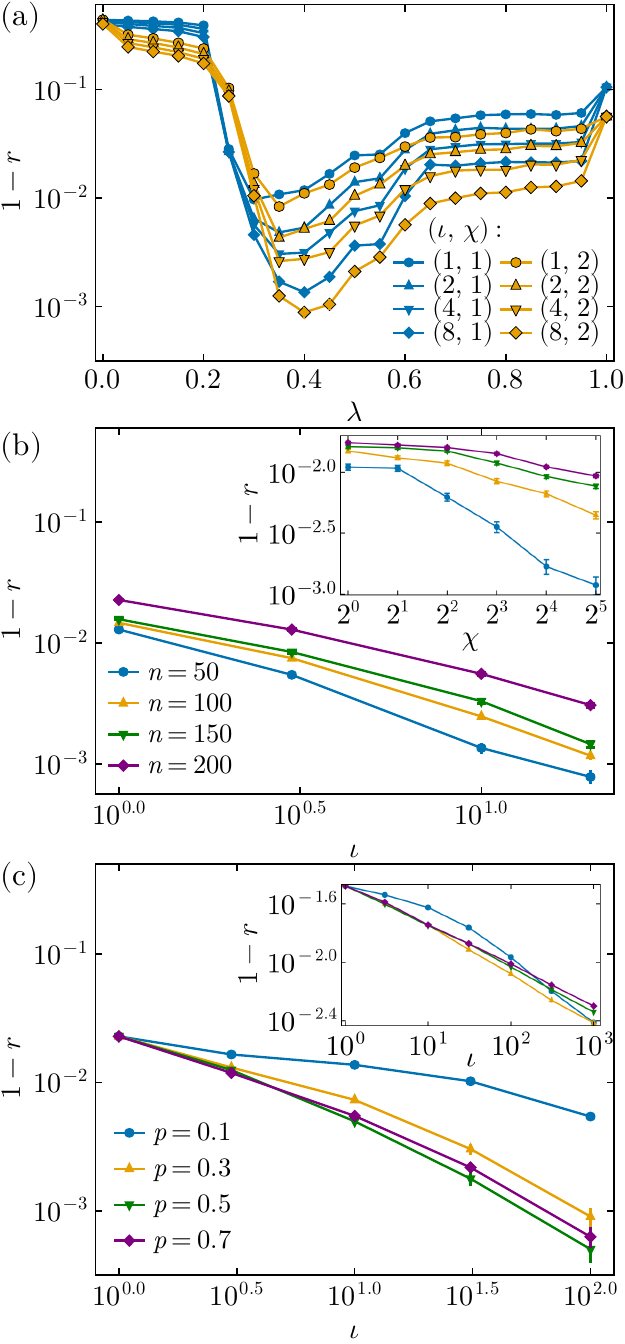}
\caption{\label{fig:2}
Study of QiILS hyperparameters.
Performance is measured by the average relative error $1 - r$, averaged over \num{1000} random graph instances.
(a) For u3R graphs of size $n = 50$, we consider various values of $\lambda$, numbers of iterations $\iota$ and bond dimensions $\chi$.
(b) For u3R graphs of different sizes, we study the performance as a function of the iteration index $\iota$ using PS ($\chi = 1$) and fixing $\lambda = 0.4$.
The inset shows the performance for the same u3R graphs when varying $\chi$ while fixing the iteration index to $\iota = 1$.
(c) For $n = 200$ and several perturbation strengths, we compare the performance for u3R graphs (main panel) and w3R graphs (inset).
[Fixed hyperparameters: sweeps = 80; in (a) and (b) $p = 0.5$; in (c) $\lambda = 0.4$ for u3R and $\lambda = 0.5$ for w3R.]
}
\end{figure}

To enable further progress and prevent the DMRG optimization from reconverging to the same local minimum, we perturb the converged MPS by randomly applying $\sigma^X$ operators to a fraction $p$ of the spins, where $p$ denotes the perturbation strength.
This is realized by randomly selecting a fraction $p$ of the MPS tensors one after another and, for each selected tensor at site $j$, by assigning
\begin{equation}
 A_{j}^{s_{j}} = A_{j}^{\neg s_{j}},
\end{equation}
where $\neg 0 = 1$ and $\neg 1 = 0$.
The perturbed state then serves as the initial condition for iteration $\iota+1$, and the process is repeated.
A more detailed description of the MPS algorithms is given in Appendix~\ref{app:QiILSUsingMPS}.

In all figures, we report the best energy obtained up to iteration $\iota$.

In Fig.~\ref{fig:2}(a), we assess the performance of QiILS as a function of $\lambda$, using a fixed perturbation value and the two bond dimensions $\chi = 1$ and $2$.
For nearly all values of $\lambda$, the performance systematically improves with both the iteration index $\iota$ and the bond dimension $\chi$.
Significant improvements over the traditional ILS baseline $(\lambda = 1 = \chi)$ appear when $\lambda \neq 1$.

In Fig.~\ref{fig:2}(b), we compare the performance scaling with the number of iterations and with the bond dimension.
For smaller problems ($n = 50$), both approaches yield comparable performance, with relative errors around $10^{-2.5}$.
However, as the problem size increases, their behavior diverges.
For larger instances ($n = 200$), the iterative method ($\chi = 1$) achieves relative errors below $10^{-2}$ that decrease systematically with $\iota$, whereas increasing the bond dimension (with only one iteration) performs worse.
Since the bond dimension dominates the computational cost of a DMRG sweep, scaling as $O(\chi^3)$ (see Appendix~\ref{app:QiILSUsingMPS}), and its benefits diminish with increasing problem size, an iterative procedure with $\chi = 1$ offers a more efficient strategy for larger MaxCut instances.

While u3R graphs are analyzed in Figs.~\ref{fig:2}(a) and (b), analogous results for w3R graphs are provided in Appendix~\ref{app:AdditionalResultsForW3RGraphs}, leading to similar conclusions.

The performance of QiILS is highly sensitive to the perturbation level.
A very small perturbation may prevent the state from escaping a local minimum, while an excessively large one effectively randomizes the configuration and has a high probability of undoing any progress made toward the ground state.
As shown in Fig.~\ref{fig:2}(c), for unweighted graphs, a perturbation strength of $p = 0.5$ yields the best performance.
For the weighted graphs studied, moderate perturbations perform better in the short term --- with $p = 0.3$ giving the lowest error for a small number of iterations --- whereas smaller perturbations ($p = 0.1$) exhibit better scaling and ultimately achieve the best performance as the number of iterations increases.

\section{Quantum-inspired iterated local search (QiILS) for $\chi = 1$}
\label{sec:QiILSForPSs}

The strong performance of QiILS at $\chi = 1$ motivates the development of a dedicated PS version of the method, which can be implemented more efficiently.
We start with a random PS of the form $\ket{\Psi} = \ket{\psi_1} \otimes \ket{\psi_2} \otimes \cdots \otimes \ket{\psi_n}$ where
\begin{equation} \label{eq:productstate}
 \ket{\psi_j} = \cos(\theta_j)\ket{0} + \sin(\theta_j)\ket{1}, \quad \theta_j \in \left[0, \frac{\pi}{2}\right].
\end{equation}
For the energy expectation value $\bra{\Psi}H\ket{\Psi}$ as a function of $\theta_j$, assuming all other angles $\theta_k$ for $k \neq j$ are fixed, we obtain the expression
\begin{align}\label{eq:esite}
\begin{split}
 E(\theta_j) &= A \cos(2\theta_j) - B \sin(2\theta_j)\\
 &= -\sqrt{A^2 + B^2} \sin[2\theta_j - \text{atan2}(A,B)],
\end{split}
\end{align}
where $A = \lambda a$,
\begin{equation}\label{eq:a}
 a = \sum_{k<j} w_{k, j} \cos(2\theta_k) + \sum_{k>j} w_{j, k} \cos(2\theta_k),
\end{equation}
and $B = 1 - \lambda$.
Equation~\eqref{eq:esite} is minimized by
\begin{equation}\label{eq:updateangle}
 \theta^{\text{new}}_j = \frac{\pi}{4} + \frac{1}{2}\text{atan2}(A,B).
\end{equation}
We update the angles sequentially, going through each $\theta_j$ in turn.
Once all angles have been updated in this manner, we refer to the full pass as a sweep.
After each sweep, we check for convergence using the condition
\begin{equation}\label{eq:convergence}
 \max_j \left| \theta^{\text{new}}_j - \theta_j \right| < \frac{\epsilon}{n} \sum_{j=1}^n \left| \theta^{\text{new}}_j - \frac{\pi}{4} \right|
\end{equation}
where $\epsilon$ is a user-defined tolerance.
Once convergence is achieved, we obtain a bitstring by projecting the PS onto the computational basis:
\begin{equation}\label{eq:rounding}
 b_j = \text{round}\left(\frac{2\theta_j}{\pi}\right) \in \{0,1\},
\end{equation}
where the rounding function assigns $b_j = 0$ for $\theta_j \le \pi/4$ and $b_j = 1$ otherwise.
The energy associated with this configuration is evaluated with respect to the spin Hamiltonian $H_\text{f}$ and recorded if it yields the lowest value obtained so far.
In this setting, the subsequent perturbation, implemented by randomly applying $\sigma^X$ operators, corresponds to updating the angles as $\theta^{\text{new}}_j = -\theta_j + \pi/2$.

\section{Performance mechanism}
\label{sec:PerformanceMechanism}

Let us investigate the mechanism underlying the performance of QiILS.
Specifically, let us address the question of why there exists a $\lambda$ regime --- visible, e.g., in Fig.~\ref{fig:2}(a) --- where the algorithms perform best.
We show that this regime corresponds to values of $\lambda$ for which the optimization landscape contains only a small number of minima.
At the same time, a large fraction of these minima map, after rounding according to Eq.~\eqref{eq:rounding}, to the lowest-energy spin configurations.
Moreover, the basins of attraction associated with these optimal-rounding minima are large and well-conditioned.
Together, these conditions enable QiILS to rapidly converge to high-quality solutions.
In the following, we present evidence supporting the proposed mechanism.

For 50 randomly generated u3R graphs of size $n = 50$, we probe the optimization landscape as a function of $\lambda$.
To this end, we minimize the energy using SciPy's Sequential Least Squares Programming (SLSQP) optimizer.
We use $1.2 \times 10^4$ randomly initialized spin configurations as starting points.
The optimization is bounded such that all angles remain within $[0, \pi/2]$.
We employ a stopping criterion of $\textsf{ftol} = 10^{-8}$ and allow up to $\textsf{maxiter} = 5 \times 10^3$ iterations per initialization.
The ground-state energy $E^{*}$ is found using an exact solver.
We denote quantities associated with minima that, after rounding, yield $E^{*}$ with a superscript $*$.

\begin{figure}
\centering
\includegraphics[width=\columnwidth]{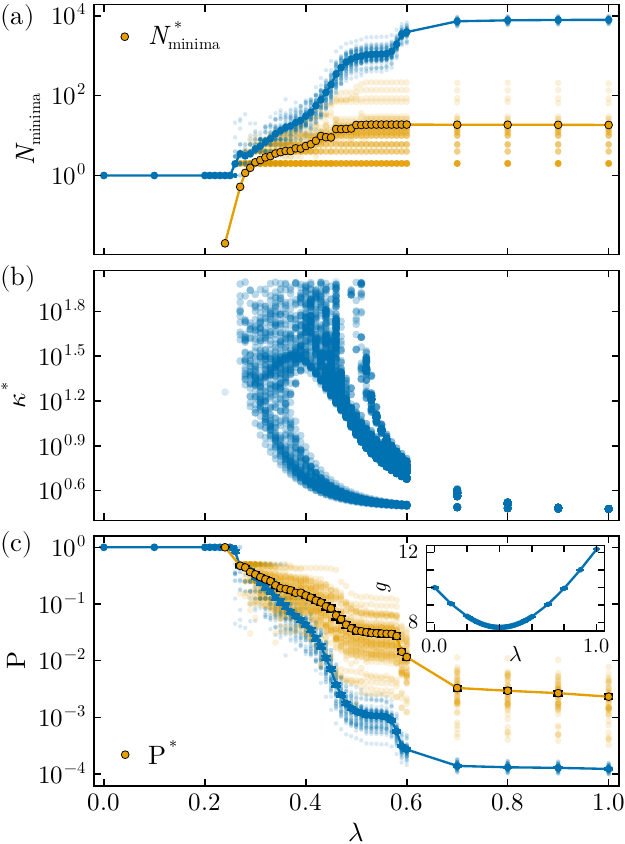}
\caption{\label{fig:3}
Optimization landscape analysis.
Shown are: (a) the total number $N_{\text{minima}}$ of distinct minima identified and the size $N^{*}_{\text{minima}}$ of the subset yielding optimal energies $E^{*}$ after rounding; (b) the condition number $\kappa^{*}$ of the landscape evaluated at the best minima, capped at a maximum value of 100; and (c) the probability that the randomly initialized SLSQP optimizer converges to each distinct minimum, where the inset shows the mean expectation value of the gradient norm, $g$.
Additional details are discussed in the main text.
}
\end{figure}

For each set of minima identified, we record the total number of distinct minima, $N_{\text{minima}}$.
We also record the size $N^{*}_{\text{minima}}$ of the subset of those distinct minima that, after rounding, yield configurations with energy $E^{*}$.
These data, including their mean values over the u3R graph instances, are shown in Fig.~\ref{fig:3}(a).
We observe that, as $\lambda$ increases, there is a transition from a single global minimum to multiple minima, many of which round to spin configurations with energy $E^{*}$.
Minima occur in symmetry-related pairs connected by the transformation $\theta_j \to -\theta_j + \pi/2$.
After the transition, $N_{\text{minima}}$ grows exponentially, whereas $N^{*}_{\text{minima}}$ increases at a much slower rate.

For each of the $N^{*}_{\text{minima}}$ distinct optimal-rounding minima, we report the condition number $\kappa^{*}$ of the corresponding Hessian matrix in Fig.~\ref{fig:3}(b).
Immediately after the transition, these minima exhibit large condition numbers.
As $\lambda$ increases further, the condition numbers decrease rapidly, such that these minima become well-conditioned in the region of $\lambda$ where QiILS and QiIGS perform best.
For the u3R graphs used in this analysis, increasing $\lambda$ leads to the appearance of many additional distinct optimal-rounding minima.
These newly emerging minima contribute to $N^{*}_{\text{minima}}$ and are initially characterized by large condition numbers that typically decrease as $\lambda$ increases.

For each distinct minimum, we also compute the probability, $P$ or $P^{*}$, that the randomly initialized optimizer converges to that minimum.
These probabilities and their mean values, averaged over all u3R graphs, are shown in Fig.~\ref{fig:3}(c).
While the mean probability across all minima is proportional to $N_{\text{minima}}^{-1}$ as expected, the values of $P^{*}$ for the optimal-rounding minima are larger on average.

The inset of Fig.~\ref{fig:3}(c) shows the mean expectation value of the gradient norm, $g$, averaged over the $50$ u3R graphs.
Here the expectation value is estimated by randomly sampling $10^4$ spin configurations.
We conclude from Fig.~\ref{fig:3}(c) that the mean gradient expectation value remains sufficiently large across all values of $\lambda$.
Therefore, the optimization landscape is navigable throughout the entire $\lambda$ range.

\section{Selection of hyperparameter $\lambda$}
\label{sec:SelectionOfHyperparameterLambda}

Ideally, one seeks to choose the value of $\lambda$ that maximizes the rate of improvement per iteration.
We propose two methods for selecting $\lambda$.
The first is to construct a diagram similar to Fig.~\ref{fig:2}(a) to visually identify the optimal value.
Alternatively, leveraging the convex profile observed in Fig.~\ref{fig:2}(a), one can determine $\lambda$ using the so-called golden-section search (GSS)~\cite{PrEtAl07}, a standard technique for locating the extremum of a function within a bounded interval.
In this approach, different $\lambda$ values are evaluated via GSS and the value that maximizes the decay rate $m$ in the fit $E(\iota) = c_0 + c_1 e^{-m \times \iota}$ is selected, where $E(\iota)$ denotes the mean energy at iteration $\iota$ averaged over multiple initial states, and $c_0$ and $c_1$ are fitting parameters.

\section{Comparison with quantum approximate optimization algorithm (QAOA)}
\label{sec:ComparisonWithQAOA}

\begin{figure}[!t]
\centering
\includegraphics[width=\columnwidth]{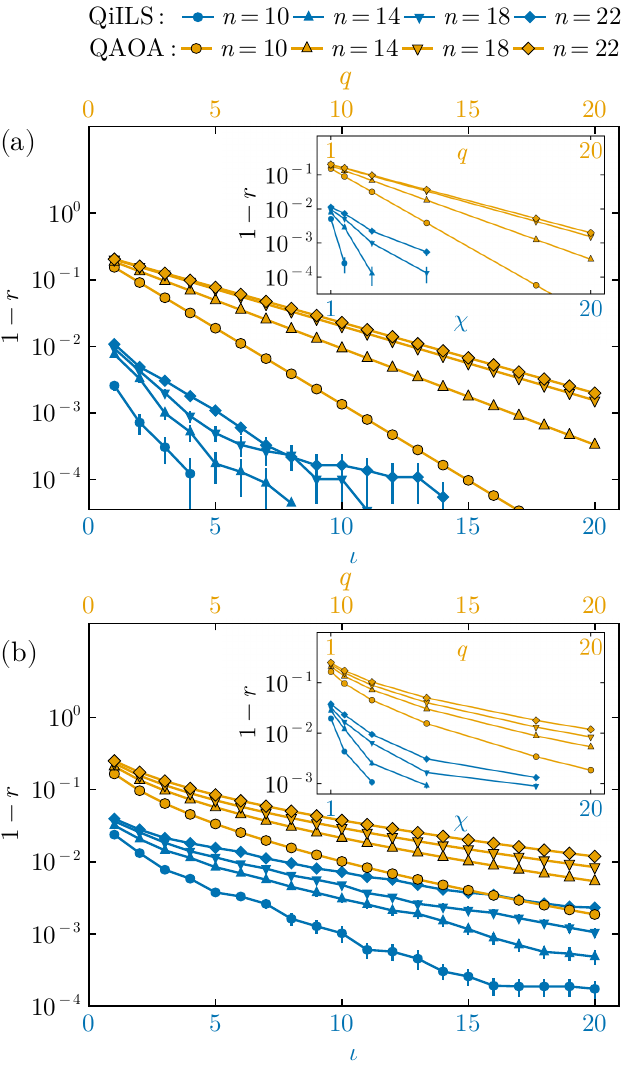}
\caption{\label{fig:4}
Comparison with QAOA.
Performance is measured by the average relative error $1 - r$, averaged over \num{1000} random graph instances.
We consider u3R graphs (a) and w3R graphs (b).
The main panels show the performance of QiILS with $\chi = 1$ as a function of $\iota$.
The insets show the performance as a function of the bond dimension $\chi$, using a single iteration ($\iota = 1$).
Closed markers are results from Fig.\ 4 of Ref.~\cite{ZhEtAl20}, where the model function is $1 - r \propto \exp(-q/q_0)$ for unweighted graphs and $1 - r \propto \exp(-\sqrt{q/q_0})$ for weighted graphs, with $q$ denoting the number of QAOA quantum circuit layers.
The quantity on the horizontal axis ($\iota$ in the main panels or $\chi$ in the insets) plays an analogous role to the QAOA circuit depth $q$.
[Fixed hyperparameters: sweeps = $80$; in (a) $\lambda = 0.55$ and $p = 0.5$; in (b) $\lambda = 0.75$ and $p = 0.3$.
Further details on the selection of $\lambda$ are provided in Appendix~\ref{app:SelectionOfLambdaForQAOAComparison}.]
}
\end{figure}

Figure~\ref{fig:4} presents a performance comparison between QiILS and QAOA, using QAOA results taken from Fig.\ 4 of Ref.~\cite{ZhEtAl20}.
We compare the performance of QiILS, evaluated either as a function of the number of iterations ($\chi = 1$) or as a function of the bond dimension (for a single iteration), to that of QAOA, whose performance is quantified by the number of QAOA quantum circuit layers.
Although these metrics are not directly comparable, they reflect analogous notions of computational effort and enable a meaningful assessment of relative efficiency.
In the unweighted case [Fig.~\ref{fig:4}(a)], we observe an exponential decay in $1 - r$ with respect to both iteration count and bond dimension.
Our method attains higher approximation ratios than those reported in Ref.~\cite{ZhEtAl20}, successfully solving all \num{1000} instances in fewer than 15 iterations.
For weighted graphs [Fig.~\ref{fig:4}(b)], QiILS also outperforms QAOA, achieving higher approximation ratios while exhibiting a comparable rate of improvement with increasing iterations.

\section{Gset benchmark}
\label{sec:GsetBenchmark}

To assess the performance of our algorithms on larger problems, we consider additional Gset problems~\cite{Gset} with $800$ variables.
As in Fig.~\ref{fig:1}, we compare the algorithms QiILS, LQA, ILS and GCS.
We benchmark them in the context of solving Gset instances G1--G10, which are based on random graph instances.
Table~\ref{tab:1} contains our results.
We observe that QiILS consistently outperforms the other solvers, in line with our conclusions drawn from Fig.~\ref{fig:1} for G12.

\begin{table*}
\centering
\begin{tabular}{|l|c|c|c|c|c|c|c|c|c|c|c|c|c|c|}
\hline
graph & solution & $\text{QiILS}_{\text{best}}$ & $\text{QiILS}_{\text{avg}}$ & $\text{QiILS}_\text{solved}$ & $\text{LQA}_{\text{best}}$ & $\text{LQA}_{\text{avg}}$ & $\text{LQA}_{\text{solved}}$ & $\text{ILS}_{\text{best}}$ & $\text{ILS}_{\text{avg}}$ & $\text{ILS}_{\text{solved}}$ & $\text{GCS}_{\text{best}}$ & $\text{GCS}_{\text{avg}}$ & $\text{GCS}_{\text{solved}}$\\
\hline
G1 & 11624 & 11624 & 11623.6 & 9/10  & 11624 & 11621.2 & 1/10 & 11622 & 11614 & 0/10 & 11624 & 11591.4 & 1/10\\
G2 & 11620 & 11620 & 11617 & 5/10  & 11608 & 11600.8 & 0/10 & 11612 & 11606.5 & 0/10 & 11600 & 11586.5 & 0/10\\
G3 & 11622 & 11622 & 11619.6 & 6/10  & 11621 & 11605.7 & 0/10 & 11615 & 11608.9 & 0/10 & 11622 & 11595.7 & 1/10\\
G4 & 11646 & 11646 & 11643.44 & 6/10  & 11646 & 11630.4 & 1/10 & 11638 & 11632 & 0/10 & 11641 & 11608.6 & 0/10\\
G5 & 11631 & 11631 & 11630.55 & 9/10 & 11623 & 11622.1 & 0/10 & 11623 & 11619.2 & 0/10 & 11622 & 11593.8 & 0/10\\
G6 & 2178 & 2178 & 2177.9 & 9/10  & 2175 & 2173 & 0/10 & 2176 & 2171.9 & 0/10 & 2160 & 2141.6 & 0/10\\
G7 & 2006 & 2006 & 2001.1 & 5/10  & 2001 & 1998 & 0/10 & 2005 & 1993.4 & 0/10 & 1994 & 1979.8 & 0/10\\
G8 & 2005 & 2005 & 2002.4 & 4/10  & 2005 & 1988.3 & 1/10 & 2001 & 1991.8 & 0/10 & 2005 & 1977.5 & 1/10\\
G9 & 2054 & 2054 & 2048.88 & 2/10 & 2045 & 2042.5 & 0/10 & 2052 & 2041 & 0/10 & 2045 & 2027.9 & 0/10\\
G10 & 2000 & 2000 & 1987.44 & 1/10 & 1993 & 1981 & 0/10 & 1997 & 1989.1 & 0/10 & 1994 & 1978.4 & 0/10\\
\hline
\end{tabular}
\caption{\label{tab:1}
Performance comparison for Gset problems~\cite{Gset}.
For each graph and algorithm, we perform ten trials, i.e.\ we run each algorithm starting from ten different initial states.
In each trial, the considered algorithm is run for \num{1000} iterations.
The subscripts ``best'', ``avg'', and ``solved'' stand for the best solution obtained, the average solution value achieved, and the fraction of times the best-known solution was obtained, respectively.
[Fixed hyperparameters: for QiILS, sweeps = 200, $p = 0.3$; for ILS, sweeps = 200, $p = 0.2$; and for LQA, $\gamma = 0.5 = \eta$.]
}
\end{table*}

\section{Quantum-inspired iterated global search (QiIGS)}
\label{sec:QiIGS}

The QiIGS algorithm replaces the sequential QiILS update of Eq.~\eqref{eq:updateangle} by a global update based on gradient descent:
\begin{equation}\label{eq:updateangle2}
 \theta^{\text{new}}_j = \theta_j - \tau \frac{\partial E}{\partial \theta_j},
\end{equation}
where $\tau$ denotes the step size,
\begin{equation}
 \frac{\partial E}{\partial \theta_j} = -2 \lambda a \sin(2\theta_j) - 2 (1 - \lambda) \cos(2\theta_j)
\end{equation}
and $a$ is defined in Eq.~\eqref{eq:a}.
This QiIGS formulation enables all angles to be updated in parallel, naturally lends itself to GPU parallelization, and therefore may be more efficient than QiILS in solving larger problem instances, involving tens of thousands of vertices or more.

\begin{figure}
\centering
\includegraphics[width=\columnwidth]{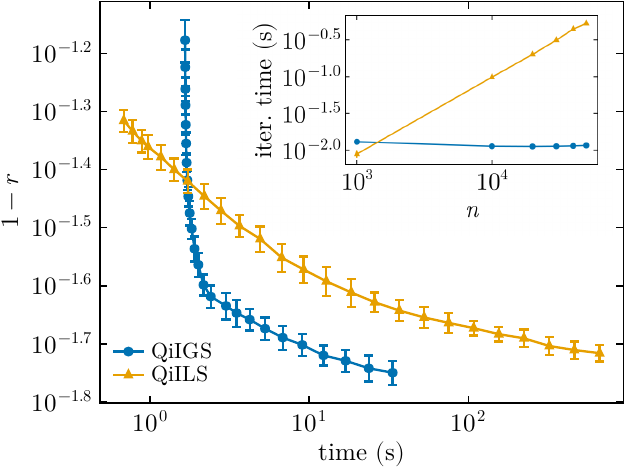}
\caption{\label{fig:5}
Comparison of QiIGS with QiILS.
In the main panel, we consider G81~\cite{Gset} and performance is measured by the average relative error $1 - r$ as a function of wall-clock time (in seconds), computed over \num{10000} iterations and averaged over 20 random initial states.
The inset shows the average time per iteration for problem sizes $n = \num{1000}, \num{10000}, \num{20000}, \num{30000}, \num{40000}, \num{50000}$ across 20 random instances of u3R graphs.
For QiIGS, angle updates were executed in parallel on an NVIDIA L4 GPU (24 GB VRAM, Ada Lovelace architecture) via Lightning AI.
The CPU computations were performed on a 2023 MacBook Pro with an Apple M2 chip (8-core CPU, 10-core GPU) and 16 GB of unified memory, running macOS 15.6.1.
[Fixed hyperparameters: sweeps = 200, $p = 0.15$, $\lambda = 0.35$, $\tau = 0.1$.]
}
\end{figure}

Indeed, Fig.~\ref{fig:5} demonstrates that for the G81 instance~\cite{Gset} --- a toroidal weighted graph with $n = \num{20000}$ vertices and weights in $\{-1, 1\}$ --- QiIGS achieves accuracy comparable to that of QiILS while offering an order-of-magnitude speed-up.
For problem sizes up to \num{50000} vertices, Fig.~\ref{fig:5} (inset) shows that GPU acceleration enables QiIGS to achieve approximately constant per-iteration times as $n$ grows.

\section{Conclusions}
\label{sec:Conclusions}

In this paper, we demonstrate that the combination of variational MPS methods --- powerful classical tools for simulating certain quantum computations --- with ILS --- an established combinatorial optimization solver --- can outperform state-of-the-art classical heuristics (ILS, LQA and GCS) and a popular variational quantum algorithm (QAOA).
From a tensor-network perspective, we find that increasing the entanglement of the optimized MPSs systematically improves the results; however, optimizing unentangled PSs over an increasing number of ILS iterations can be significantly more efficient and yield better final solutions.
From a combinatorial-optimization viewpoint, the introduced QiILS and QiIGS algorithms strictly generalize ILS, recovering it as a special case for $\lambda = 1$, while enabling parallelization strategies that fundamentally extend beyond the sequential update structure of standard ILS.
Finally, from a quantum-computing perspective, the proposed techniques are conceptually simple, naturally amenable to GPU parallelization, and therefore provide a useful classical framework to benchmark and challenge current and future quantum algorithms for combinatorial optimization.

\acknowledgments
We thank Marco Ballarin and Lewis Wright for valuable feedback on the article.
We are also grateful to Pranav Kalidindi for assistance with the GPU service.

\appendix

\section{Overview of benchmark optimization methods}
\label{app:OverviewOfBenchmarkOptimizationMethods}

Here, we provide an overview of the combinatorial optimization methods used as baselines for comparisons in the main text.
Each subsection summarizes one method, highlighting its key features and implementation details.
In the final subsection, we present hyperparameter exploration simulations used to select the settings employed in the benchmarks.

\subsection{Iterated local search (ILS)}

Iterated local search (ILS)~\cite{LoMaSt19} is a metaheuristic that repeatedly invokes a local search procedure, each time starting from a perturbed version of a previously found solution.
In our case, the local search step consists of finding the ground state of the Hamiltonian $H_\text{f}$.
Once a local minimum is reached, we apply a perturbation --- specifically, a set of $\sigma^X$ operators --- to escape the local region of convergence and continue the search.
This process is iterated until a satisfactory solution is obtained.

\subsection{Local quantum annealing (LQA)}

Local quantum annealing (LQA)~\cite{BoEtAl22} is an approach based on a parameterized product state (PS) $\ket{\Psi}$ and takes inspiration from quantum annealing.
Starting in the ground state of the initial Hamiltonian $H_\text{i} = -\sum_j \sigma_{j}^X$, LQA sequentially minimizes the energy of the PS with respect to the Hamiltonian
\begin{equation}\label{eq:hamlqa}
 H(t) = \left(1 - \frac{t}{T}\right) H_\text{i} + \frac{t}{T}\, \gamma H_\text{f},
\end{equation}
at a discrete set of times $0 < t_1 < t_2 < \ldots < t_M = T$.
In Eq.~\eqref{eq:hamlqa}, the ground state(s) of the Hamiltonian $H_\text{f}$ encode(s) the optimal solution(s) to the considered combinatorial optimization problem and the scalar $\gamma$ rescales $H_\text{f}$.
The energy minimization is performed via gradient descent with a step size $\eta$.
In particular, at each time step $t_j$, the gradients of the energy expectation value $\bra{\Psi} H(t_{j}) \ket{\Psi}$ are evaluated and the parameters of the PS are updated once in the direction of steepest descent.

\subsection{Generalized group-theoretic coherent states (GCS)}

\begin{figure*}
\centering
\includegraphics[width=\textwidth]{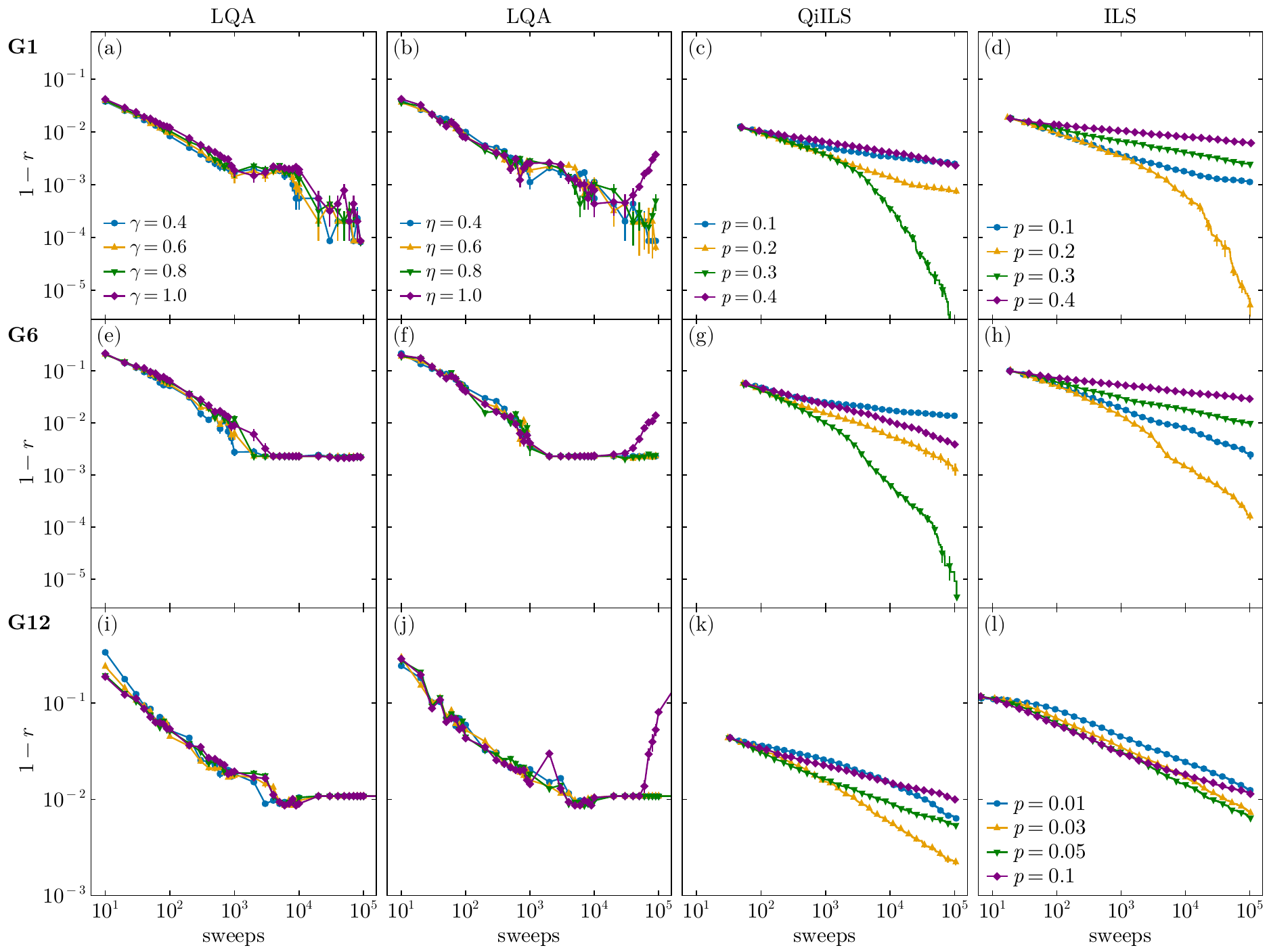}
\caption{\label{fig:6}
Hyperparameter exploration for LQA, QiILS, and ILS applied to G1 (a--d), G6 (e--h), and G12 (i--l)~\cite{Gset}.
Average performance is measured by the average relative error $1-r$ and various hyperparameter choices are considered.
(a, e, i) In LQA, we fix $\eta = 0.5$, vary $\gamma \in \{0.4, 0.6, 0.8, 1.0\}$, and average over 20 (a, e) or 10 (i) random initial states.
(b, f, j) In LQA, we fix $\gamma = 0.5$, vary $\eta \in \{0.4, 0.6, 0.8, 1.0\}$, and average over 20 (b, f) or 10 (j) random initial states.
(c, g, k) In QiILS, we vary the perturbation strength $p \in \{0.1, 0.2, 0.3, 0.4\}$, use $\lambda = 0.4$ (c) or $\lambda = 0.3$ (g, k), and average over 100 (c, g) or 20 (k) random initial states.
(d, h, l) In ILS, we vary the perturbation strength $p \in \{0.1, 0.2, 0.3, 0.4\}$ (d, h) or $p \in \{0.01, 0.03, 0.05, 0.1\}$ (l), and average over 100 (d, h) or 20 (l) random initial states.
}
\end{figure*}

The algorithm of Ref.~\cite{FiSa25}, based on generalized group-theoretic coherent states (GCS)~\cite{GuEtAl21, ScEtAl22}, is a quantum-inspired variational method for combinatorial optimization.
Similar to LQA, it employs a parameterized ansatz to mimic the time-dependent evolution of quantum annealing, but within a more expressive variational manifold.
Starting with the PS $\ket{+}^{\otimes n} = [(\ket{0}+\ket{1})/\sqrt{2}]^{\otimes n}$, the GCS ansatz is constructed by applying two types of unitaries:
single-qubit rotations
\begin{equation}
 \mathcal{U}(x) = \bigotimes_{j=1}^n \exp(-\text{i} \sum_{K \in \{X,Y,Z\}} x_{j, K}\, \sigma^{K}_j)
\end{equation}
and two-qubit rotations
\begin{equation}
 \mathcal{V}(B) = \exp(-\text{i} \sum_{j \neq k} B_{j, k}\, \sigma^{Z}_j \sigma^{Z}_k).
\end{equation}
These operations are combined to produce the state
\begin{equation}
 \ket{\Psi(x,B,y)} = \mathcal{U}(y)\,\mathcal{V}(B)\,\mathcal{U}(x)\ket{+}^{\otimes n}.
\end{equation}
The variational parameters consist of two real $n \times 3$ matrices $x$ and $y$, and a symmetric $n \times n$ matrix $B$.
Setting all variational parameters equal to 0 initializes the state in $\ket{+}^{\otimes n}$.
To emulate quantum annealing, we use the interpolating Hamiltonian~\eqref{eq:hamlqa} of LQA, however, without the rescaling paramater $\gamma$.
As in LQA, the total evolution time $T$ is discretized into points $0 < t_1 < t_2 < \ldots < t_M = T$.  
At each time $t_j$, the parameters $(x,B,y)$ are updated via one steepest descent step using the gradient of $\bra{\Psi(x,B,y)} H(t_j) \ket{\Psi(x,B,y)}$.

\subsection{Hyperparameter exploration}

In the following, we discuss the results for different settings of the hyperparameters for the methods LQA, QiILS and ILS.
Experiments are conducted on Gset instances G1, G6 and G12~\cite{Gset} and the results are shown in Fig.~\ref{fig:6}.

\begin{table*}
\centering
\begin{tabular}{|l|c|c|c|c|}
\hline
 & $p$ & sweeps & iterations & $\lambda$\\
\hline
Fig.~\ref{fig:1} (QiILS) & 0.2 & 200 & -- & --\\
Fig.~\ref{fig:1} (ILS) & 0.03 & 200 & -- & 1.0\\
Fig.~\ref{fig:2}(a) & 0.5 & 80 & 1,2,4,8 & --\\
Fig.~\ref{fig:2}(b) & 0.5 & 80 & -- & 0.4\\
Fig.~\ref{fig:2}(b) (inset) & -- & 80 & 1 & 0.4\\
Fig.~\ref{fig:2}(c) & 0.1,0.3,0.5,0.7 & 80 & -- & 0.4\\
Fig.~\ref{fig:2}(c) (inset) & 0.1,0.3,0.5,0.7 & 80 & -- & 0.5\\
Fig.~\ref{fig:4}(a) & 0.5 & 80 & -- & 0.55\\
Fig.~\ref{fig:4}(a) (inset) & -- & 80 & 1 & 0.55\\
Fig.~\ref{fig:4}(b) & 0.3 & 80 & -- & 0.75\\
Fig.~\ref{fig:4}(b) (inset) & -- & 80 & 1 & 0.75\\
Fig.~\ref{fig:5} & 0.15 & 200 & \num{10000} & 0.35\\
Fig.~\ref{fig:5} (inset) & -- & 200 & 1 & 0.35\\
Fig.~\ref{fig:6}(c) & 0.1,0.2,0.3,0.4 & -- & -- & 0.4\\
Fig.~\ref{fig:6}(d) & 0.1,0.2,0.3,0.4 & -- & -- & 1.0\\
Fig.~\ref{fig:6}(g) & 0.1,0.2,0.3,0.4 & -- & -- & 0.3\\
Fig.~\ref{fig:6}(h) & 0.1,0.2,0.3,0.4 & -- & -- & 1.0\\
Fig.~\ref{fig:6}(k) & 0.1,0.2,0.3,0.4 & -- & -- & 0.3\\
Fig.~\ref{fig:6}(l) & 0.01,0.03,0.05,0.1 & -- & -- & 1.0\\
Fig.~\ref{fig:8}(a) & 0.3 & 80 & 1,2,4,8  & --\\
Fig.~\ref{fig:8}(b) & 0.3 & 80 & -- & 0.6\\
Fig.~\ref{fig:8}(b) (inset) & -- & 80 & 1 & 0.6\\
Fig.~\ref{fig:9}(a) & 0.5 & 80 & 1,2,4 & --\\
Fig.~\ref{fig:9}(b) & 0.3 & 80 & 1,2,4,8 & --\\
\hline
G1 & 0.3 & 200 & \num{1000} & 0.38\\
G2 & 0.3 & 200 & \num{1000} & 0.41\\
G3 & 0.3 & 200 & \num{1000} & 0.38\\
G4 & 0.3 & 200 & \num{1000} & 0.38\\
G5 & 0.3 & 200 & \num{1000} & 0.3\\
G6 & 0.2 & 200 & \num{1000} & 0.42\\
G7 & 0.2 & 200 & \num{1000} & 0.41\\
G8 & 0.2 & 200 & \num{1000} & 0.38\\
G9 & 0.2 & 200 & \num{1000} & 0.41\\
G10 & 0.2 & 200 & \num{1000} & 0.38\\
G12 & 0.2 & 200 & -- & --\\
G81 & 0.15 & 200 & \num{10000} & 0.35\\
\hline
\end{tabular}
\caption{\label{tab:2}
Numerical values of perturbation, sweeps, iterations and $\lambda$ used for all figures and tables of this paper.
}
\end{table*}

For LQA, we focus on the hyperparameters $\gamma$ and $\eta$, which control the rescaling of the problem Hamiltonian and the gradient descent step size, respectively.
For each pair of hyperparameters, we fix one while sweeping over a range of values for the other to assess its influence on performance.
Across the tested range, the hyperparameter $\gamma$ does not lead to significant variations in performance.
In contrast, the step size $\eta$ has a noticeable effect:
Large values (e.g., $\eta = 1.0$) can cause the method to diverge from the solution when the number of sweeps is large, as illustrated in Fig.~\ref{fig:6}(b, f, j).
For G1, we conclude from Fig.~\ref{fig:6}(a, b) that all studied hyperparameter choices yield similar systematic improvements of the results with increasing number of sweeps, except for $\eta = 0.8$ and $1.0$ where larger numbers of sweeps can lead to worse results.
For G6 and G12, we conclude from Fig.~\ref{fig:6}(e, f, i, j) that across all hyperparameter settings the results plateau ($\eta < 1.0$) or deteriorate ($\eta = 1.0$) for larger numbers of sweeps.

For QiILS and ILS, we vary the perturbation strength $p$ over a range of values.
In both methods, the perturbation strength has a significant impact on performance, as shown in Fig.~\ref{fig:6}(c, d, g, h, k, l).
We observe that achieving the best performance requires smaller perturbations in ILS than in QiILS, indicating that ILS is more sensitive to strong perturbations.

\section{Hyperparameters used}
\label{app:HyperparametersUsed}

Table~\ref{tab:2} summarizes the hyperparameters employed in all numerical simulations presented in this paper.
Note that the value of sweeps in Tab.~\ref{tab:2} is an upper bound, as we stop an iteration when the optimization has converged.
Also, in Fig.~\ref{fig:1}, to find $\lambda$, we employ 50 random initializations, 25 iterations, and a maximum of 200 sweeps per iteration.
For the LQA results of Fig.~\ref{fig:1} and related to the Gset instances, we use $\gamma = 0.5 = \eta$.
To find the optimal $\lambda$ values for the Gset calculations, 10 random initializations and 50 iterations were used for each graph.

\begin{table*}
\centering
\begin{tabular}{|l|c|c|c|l|}
\hline
algorithm & median (ms) & IQR (ms) & relative to QiILS & remarks\\
\hline
QiILS & 0.05 & 0.001 & $\times$1.0 & baseline (iterative local sweeps)\\
ILS & 0.05 & 0.001 & $\times$1.0 & same per-sweep update cost as QiILS\\
LQA & 0.34 & 0.00  & $\times$6.8 & PyTorch-based; heavier local update rule\\
GCS & 463.18 & 11.93 & $\times$\num{9260} & variational updates; large matrix overhead\\
\hline
\end{tabular}
\caption{\label{tab:3}
Median CPU time in milliseconds (ms) per one sweep over all variables for various algorithms solving the G12 problem~\cite{Gset}, where the median is calculated across 100 runs of each algorithm starting from different random initializations.
The ``median'' indicates the typical time required per sweep, while the ``IQR (interquartile range)'' represents the variability between the 25th and 75th percentiles across runs.
All experiments were performed on an Apple M2 CPU (8-core, 16 GB RAM) without GPU acceleration.
}
\end{table*}

\section{Times per sweep for each method}
\label{app:TimesPerSweepForEachMethod}

Table~\ref{tab:3} reports the median wall-clock time per sweep for each optimization method considered in Fig.~\ref{fig:1}.
All timings were obtained under identical hardware and software conditions.
We can see in Tab.~\ref{tab:3} that QiILS and ILS exhibit identical per-sweep efficiency, whereas LQA and GCS show progressively higher computational overhead due to their more complex update mechanisms.

\section{Quantum-inspired iterated local search (QiILS) using matrix product states (MPS)}
\label{app:QiILSUsingMPS}

Here, we describe the QiILS algorithm based on MPS in more detail than is provided in the main text.
In particular, we go beyond the main text by explaining MPS basics and the established MPS algorithms used in QiILS.
We note that these topics are covered in review articles~\cite{VeMuCi08, Sc11, Or14, Ba23}, which the reader can consult for further details.

The MPS-based QiILS procedure starts from a randomly initialized MPS and iterates the following three steps:
\begin{enumerate}
\item The expectation value of the quantum annealing Hamiltonian is minimized using the MPS formulation of the density matrix renormalization group (DMRG).
\item From the resulting MPS, a bitstring is extracted as a candidate solution for the combinatorial optimization problem.
\item The MPS obtained after the DMRG optimization is perturbed by applying Pauli-$X$ matrices, after which the procedure returns to step 1.
\end{enumerate}

\begin{figure*}
\centering
\includegraphics[width=\textwidth]{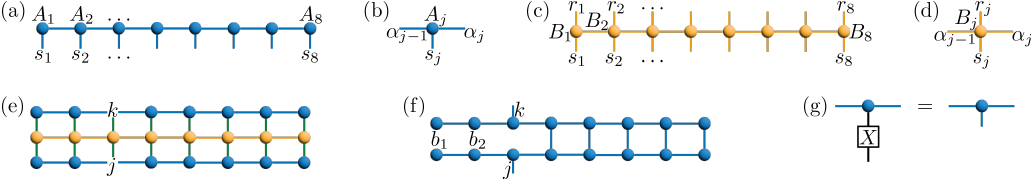}
\caption{\label{fig:7}
Graphical tensor-network representations.
Tensors are depicted as geometric shapes (spheres and rectangles) and their indices are represented by edges.
Connecting two tensors by an edge indicates summation over the corresponding index.
An open edge --- attached to a tensor but not connected to another --- represents an uncontracted degree of freedom.
Shown are: (a) an MPS for $n = 8$ qubits; (b) an internal MPS tensor $A_j$ for $1 < j < n$; (c) an MPO for $n = 8$ qubits; (d) an internal MPO tensor $B_j$ for $1 < j < n$; (e) the effective Hamiltonian $H_{j, k}^{\text{eff}}$ at site 3, assuming the MPO represents the Hamiltonian; (f) the reduced density matrix $\rho_{j, k}^{(3)}$, obtained after projecting the first two physical indices onto $b_1$ and $b_2$, respectively; and (g) the application of a Pauli-$X$ matrix to an MPS tensor.
}
\end{figure*}

Let us first explain the MPS and matrix-product-operator (MPO) ans\"{a}tze.
We assume open boundary conditions.
An MPS of $n$ qubits is given by
\begin{equation}\label{eq:MPS}
 \ket{\Psi} = \sum_{\{s_j, \alpha_j\}} A_{1, \alpha_1}^{s_1} A_{2, \alpha_1, \alpha_2}^{s_2} \ldots A_{n, \alpha_{n-1}}^{s_n} \ket{s_1, s_2, \ldots, s_n}
\end{equation}
where $s_j \in \{0, 1\}$ and $\alpha_j \in \{1, 2, \ldots, \chi\}$.
An MPO has the form
\begin{equation}\label{eq:MPO}
 \mathcal{O} = \sum_{\{s_j, r_j, \alpha_j\}} B_{1, \alpha_1}^{s_1, r_1} B_{2, \alpha_1, \alpha_2}^{s_2, r_2} \ldots \ketbra{s_1, \ldots, s_n}{r_1, \ldots, r_n}
\end{equation}
where $r_j \in \{0, 1\}$.
The MPS (MPO) ansatz consists of a product of $n$ tensors, contracted via their so-called virtual indices $\{\alpha_j\}$.
The indices $\{r_j, s_j\}$ are referred to as physical indices, as they correspond to the physical degrees of freedom of the qubits.
The so-called bond dimension $\chi$ determines the expressivity of the ansatz as well as the computational cost of working with it.
In practice, explicit mathematical equations are rarely used in the MPS and MPO formalism, as keeping track of the many indices is cumbersome.
Instead, one typically adopts a graphical representation, illustrated in Fig.~\ref{fig:7}(a--d) for an MPS and MPO.

The MPS variant of DMRG seeks to minimize the energy expectation value
\begin{equation}
 E = \frac{\bra{\Psi}\!H\!\ket{\Psi}}{\braket{\Psi}{\Psi}}
\end{equation}
of a Hamiltonian $H$ by variationally optimizing the MPS parameters.
We consider the quantum annealing Hamiltonian of Eq.~\eqref{eq:adiab} and make use of its MPO representation, for which an analytical expression exists with minimum bond dimension $\tilde{\chi} \approx n/2$~\cite{FrNeDu10}.
The DMRG approach performs sweeps over the MPS tensors and for each tensor determines the optimal entries that minimize $E$, while keeping all other tensors fixed.
The optimal tensor elements are given by the eigenvector associated with the lowest eigenvalue of an effective Hamiltonian.
This effective Hamiltonian is obtained from the tensor network corresponding to $\bra{\Psi}\!H\!\ket{\Psi}$ by removing the considered tensor and its adjoint.
Figure~\ref{fig:7} (e) shows the graphical notation for the effective Hamiltonian associated with the third MPS tensor.
In practice, the effective Hamiltonian is not constructed explicitly; instead, only its action on a vector is evaluated.
As a result, the computational cost per sweep of the DMRG method scales as $O(n \chi^3 \tilde{\chi}) + O(n \chi^2 \tilde{\chi}^2)$.

After convergence of the DMRG energy minimization, we extract a bitstring $(b_1, b_2, \ldots, b_n)$ from the resulting MPS by choosing the most likely values of the physical MPS indices one after another.
We begin with the first MPS site and compute the corresponding reduced density matrix
\begin{equation}
 \rho^{(1)} = \text{tr}_{\{s_2, \ldots, s_n\}} \left( \ketbra{\Psi}{\Psi} \right),
\end{equation}
where $\text{tr}_S(\cdot)$ denotes the partial trace over the index set $S$.
We assign to $b_1$ the bit value $0$ or $1$ maximizing $\bra{b_1}\!\rho^{(1)}\!\ket{b_1}$.
Next, we fix the first physical MPS index to $b_1$ and compute the reduced density matrix for the second MPS site,
\begin{equation}
 \rho^{(2)} = \text{tr}_{\{s_3, \ldots, s_n\}} \left[ \left(\bra{b_1} \otimes \mathds{1}\right) \ketbra{\Psi}{\Psi} \left(\ket{b_1} \otimes \mathds{1}\right) \right],
\end{equation}
where $\bra{b_1}$ and $\ket{b_1}$ act on the first MPS site and the identity $\mathds{1}$ acts on the remaining MPS sites.
The bit value $b_2$ is chosen as the value maximizing $\bra{b_2}\!\rho^{(2)}\!\ket{b_2}$.
We continue by fixing the first and second physical MPS indices to $b_1$ and $b_2$, respectively, and computing the reduced density matrix for the third MPS site,
\begin{equation}
 \rho^{(3)} = \text{tr}_{\{s_4, \ldots, s_n\}} \left[ \left(\bra{b_1, b_2} \otimes \mathds{1}\right) \ketbra{\Psi}{\Psi} \left(\ket{b_1, b_2} \otimes \mathds{1}\right) \right].
\end{equation}
Figure~\ref{fig:7}(f) depicts the tensor-network representation of $\rho^{(3)}$.
We choose the most likely bit value for $b_3$ and continue the scheme until all $\{b_j\}$ have been assigned.
Overall, the computational cost of this approach scales as $O(n \chi^3)$.
The resulting bitstring $(b_1, b_2, \ldots, b_n)$ serves as a candidate solution for the combinatorial optimization problem.
We compute its energy with respect to the final Hamiltonian and keep track of the lowest energy achieved.

Having obtained a bitstring, we next perturb the previously converged MPS.
We sequentially and randomly select a total of $p n$ MPS tensors, where $p$ denotes the perturbation strength.
Each selected tensor is multiplied via its physical index with a Pauli-$X$ matrix.
The total computational cost of the perturbation step scales as $O(p n \chi^2)$.
An example of a perturbation application is shown in Fig.~\ref{fig:7}(g).
The perturbed MPS serves as the initial state for the subsequent DMRG run.

In MPS-based QiILS, this process --- DMRG sweeps, bitstring extraction, perturbative updates --- is iterated.
The dominant computational cost arises from the DMRG sweeps and scales as $O(n^2 \chi^3) + O(n^3 \chi^2)$ per sweep.
For fixed $n$, the leading asymptotic scaling is $O(\chi^3)$.

\section{Additional results for w3R graphs}
\label{app:AdditionalResultsForW3RGraphs}

\begin{figure*}
\centering
\includegraphics[width=\textwidth]{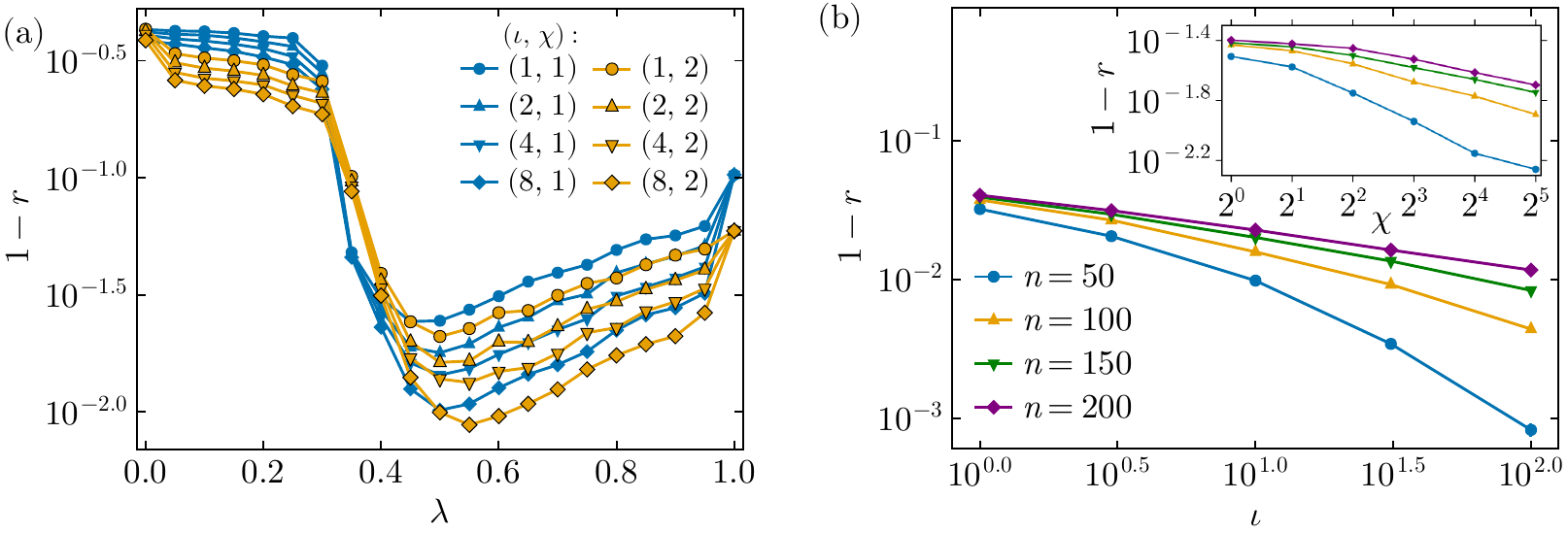}
\caption{\label{fig:8}
Study of QiILS hyperparameters for w3R graphs.
Performance is measured by the average relative error $1 - r$, averaged over 500 random w3R graph instances.
(a) For $n = 50$, we consider several values of $\lambda$, numbers of iterations $\iota$ and bond dimensions $\chi$.
(b) For graphs of various sizes, we investigate the performance as a function of the iteration index $\iota$ using $\chi = 1$ at $\lambda = 0.6$.
The inset shows the performance for the same graphs as a function of $\chi$ and fixing $\iota = 1$.
[Fixed hyperparameters: sweeps = 80, $p = 0.3$.]
}
\end{figure*}

Here, we evaluate QiILS's performance across w3R graphs of sizes $n = 50, 100, 150$, and $200$.
These graphs are the weighted counterparts of the u3R graphs considered in Figs.~\ref{fig:2}(a) and (b).
Figure~\ref{fig:8} shows our results.
We can see in Fig~\ref{fig:8}(a) that the improvement achieved for $\lambda \neq 1$ over $\lambda = 1$ is not as large as for our unweighted graph results of Fig.~\ref{fig:2}(a).
However, the characteristic convex profile with a well-defined optimal $\lambda$ region is clearly visible, as in Fig.~\ref{fig:2}(a).
Figure~\ref{fig:8}(b) shows that increasing the number of iterations with a PS ansatz performs better than using a larger bond dimension, similar to what we can see in Fig.~\ref{fig:2}(b).

\section{Selection of $\lambda$ for QAOA comparison}
\label{app:SelectionOfLambdaForQAOAComparison}

\begin{figure*}
\centering
\includegraphics[width=\textwidth]{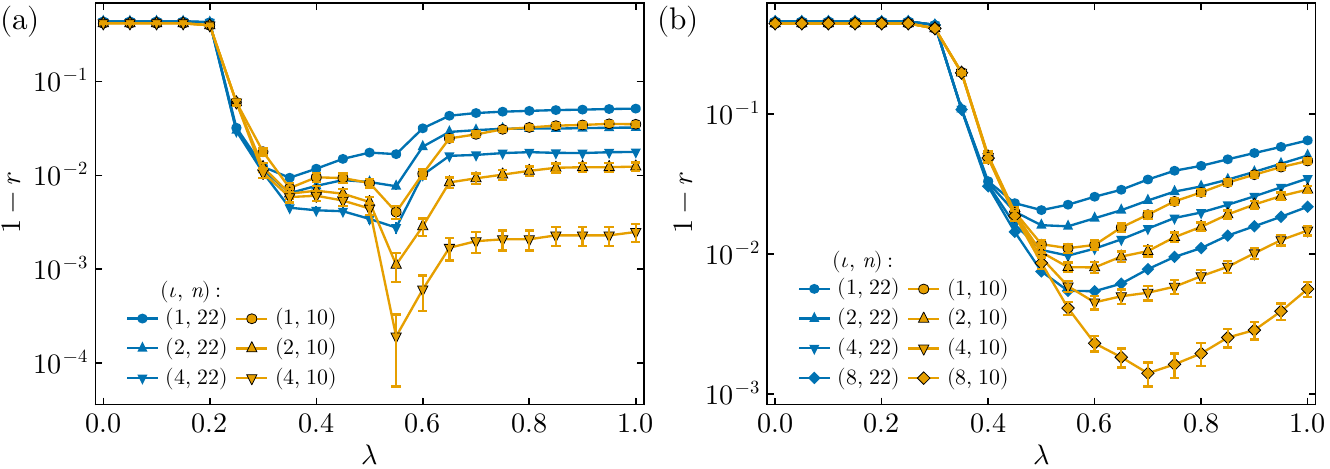}
\caption{\label{fig:9}
Selection of $\lambda$.
Performance is measured by the average relative error $1 - r$, averaged over 800 random instances, for different values of $\lambda$.
(a) For u3R graphs, we find that the optimal value for $\lambda$ is $\approx 0.55$.
(b) For w3R graphs, the optimal value of $\lambda$ is $\approx 0.75$.
[Fixed hyperparameters: sweeps = 80; in (a) $p = 0.5$; in (b) $p = 0.3$.]
}
\end{figure*}

To select the hyperparameter $\lambda$ used in Fig.~\ref{fig:4}, we perform an analysis analogous to the one related to the diagram shown in Fig.~\ref{fig:2}(a).
Such diagrams provide a practical guide for tuning $\lambda$ for specific problem instances.
Our results are shown in Fig.~\ref{fig:9}.

\newpage

\bibliography{references.bib}

\begin{thebibliography}{69}%
\makeatletter
\providecommand \@ifxundefined [1]{%
 \@ifx{#1\undefined}
}%
\providecommand \@ifnum [1]{%
 \ifnum #1\expandafter \@firstoftwo
 \else \expandafter \@secondoftwo
 \fi
}%
\providecommand \@ifx [1]{%
 \ifx #1\expandafter \@firstoftwo
 \else \expandafter \@secondoftwo
 \fi
}%
\providecommand \natexlab [1]{#1}%
\providecommand \enquote  [1]{``#1''}%
\providecommand \bibnamefont  [1]{#1}%
\providecommand \bibfnamefont [1]{#1}%
\providecommand \citenamefont [1]{#1}%
\providecommand \href@noop [0]{\@secondoftwo}%
\providecommand \href [0]{\begingroup \@sanitize@url \@href}%
\providecommand \@href[1]{\@@startlink{#1}\@@href}%
\providecommand \@@href[1]{\endgroup#1\@@endlink}%
\providecommand \@sanitize@url [0]{\catcode `\\12\catcode `\$12\catcode
  `\&12\catcode `\#12\catcode `\^12\catcode `\_12\catcode `\%12\relax}%
\providecommand \@@startlink[1]{}%
\providecommand \@@endlink[0]{}%
\providecommand \url  [0]{\begingroup\@sanitize@url \@url }%
\providecommand \@url [1]{\endgroup\@href {#1}{\urlprefix }}%
\providecommand \urlprefix  [0]{URL }%
\providecommand \Eprint [0]{\href }%
\providecommand \doibase [0]{https://doi.org/}%
\providecommand \selectlanguage [0]{\@gobble}%
\providecommand \bibinfo  [0]{\@secondoftwo}%
\providecommand \bibfield  [0]{\@secondoftwo}%
\providecommand \translation [1]{[#1]}%
\providecommand \BibitemOpen [0]{}%
\providecommand \bibitemStop [0]{}%
\providecommand \bibitemNoStop [0]{.\EOS\space}%
\providecommand \EOS [0]{\spacefactor3000\relax}%
\providecommand \BibitemShut  [1]{\csname bibitem#1\endcsname}%
\let\auto@bib@innerbib\@empty
\bibitem [{\citenamefont {Zhang}\ \emph {et~al.}(2023)\citenamefont {Zhang},
  \citenamefont {Wu}, \citenamefont {Ma}, \citenamefont {Song}, \citenamefont
  {Le}, \citenamefont {Cao},\ and\ \citenamefont {Zhang}}]{ZhEtAl23}%
  \BibitemOpen
  \bibfield  {author} {\bibinfo {author} {\bibfnamefont {C.}~\bibnamefont
  {Zhang}}, \bibinfo {author} {\bibfnamefont {Y.}~\bibnamefont {Wu}}, \bibinfo
  {author} {\bibfnamefont {Y.}~\bibnamefont {Ma}}, \bibinfo {author}
  {\bibfnamefont {W.}~\bibnamefont {Song}}, \bibinfo {author} {\bibfnamefont
  {Z.}~\bibnamefont {Le}}, \bibinfo {author} {\bibfnamefont {Z.}~\bibnamefont
  {Cao}},\ and\ \bibinfo {author} {\bibfnamefont {J.}~\bibnamefont {Zhang}},\
  }\bibfield  {title} {\bibinfo {title} {A review on learning to solve
  combinatorial optimisation problems in manufacturing},\ }\href
  {https://doi.org/https://doi.org/10.1049/cim2.12072} {\bibfield  {journal}
  {\bibinfo  {journal} {IET Collab. Intell. Manuf.}\ }\textbf {\bibinfo
  {volume} {5}},\ \bibinfo {pages} {e12072} (\bibinfo {year}
  {2023})}\BibitemShut {NoStop}%
\bibitem [{\citenamefont {Guihaire}\ and\ \citenamefont {Hao}(2008)}]{GuHa08}%
  \BibitemOpen
  \bibfield  {author} {\bibinfo {author} {\bibfnamefont {V.}~\bibnamefont
  {Guihaire}}\ and\ \bibinfo {author} {\bibfnamefont {J.-K.}\ \bibnamefont
  {Hao}},\ }\bibfield  {title} {\bibinfo {title} {{Transit network design and
  scheduling: A global review}},\ }\href
  {https://doi.org/https://doi.org/10.1016/j.tra.2008.03.011} {\bibfield
  {journal} {\bibinfo  {journal} {Transp. Res. A}\ }\textbf {\bibinfo {volume}
  {42}},\ \bibinfo {pages} {1251} (\bibinfo {year} {2008})}\BibitemShut
  {NoStop}%
\bibitem [{\citenamefont {Farahani}\ \emph {et~al.}(2013)\citenamefont
  {Farahani}, \citenamefont {Miandoabchi}, \citenamefont {Szeto},\ and\
  \citenamefont {Rashidi}}]{FaEtAl13}%
  \BibitemOpen
  \bibfield  {author} {\bibinfo {author} {\bibfnamefont {R.~Z.}\ \bibnamefont
  {Farahani}}, \bibinfo {author} {\bibfnamefont {E.}~\bibnamefont
  {Miandoabchi}}, \bibinfo {author} {\bibfnamefont {W.~Y.}\ \bibnamefont
  {Szeto}},\ and\ \bibinfo {author} {\bibfnamefont {H.}~\bibnamefont
  {Rashidi}},\ }\bibfield  {title} {\bibinfo {title} {A review of urban
  transportation network design problems},\ }\href
  {https://doi.org/https://doi.org/10.1016/j.ejor.2013.01.001} {\bibfield
  {journal} {\bibinfo  {journal} {Eur. J. Oper. Res.}\ }\textbf {\bibinfo
  {volume} {229}},\ \bibinfo {pages} {281} (\bibinfo {year}
  {2013})}\BibitemShut {NoStop}%
\bibitem [{\citenamefont {Liu}\ \emph {et~al.}(2024)\citenamefont {Liu},
  \citenamefont {Chang}, \citenamefont {Hong}, \citenamefont {Wu},
  \citenamefont {Man-Cho~So}, \citenamefont {Jorswieck},\ and\ \citenamefont
  {Yu}}]{LiEtAl24}%
  \BibitemOpen
  \bibfield  {author} {\bibinfo {author} {\bibfnamefont {Y.-F.}\ \bibnamefont
  {Liu}}, \bibinfo {author} {\bibfnamefont {T.-H.}\ \bibnamefont {Chang}},
  \bibinfo {author} {\bibfnamefont {M.}~\bibnamefont {Hong}}, \bibinfo {author}
  {\bibfnamefont {Z.}~\bibnamefont {Wu}}, \bibinfo {author} {\bibfnamefont
  {A.}~\bibnamefont {Man-Cho~So}}, \bibinfo {author} {\bibfnamefont {E.~A.}\
  \bibnamefont {Jorswieck}},\ and\ \bibinfo {author} {\bibfnamefont
  {W.}~\bibnamefont {Yu}},\ }\bibfield  {title} {\bibinfo {title} {{A Survey of
  Recent Advances in Optimization Methods for Wireless Communications}},\
  }\href {https://doi.org/10.1109/JSAC.2024.3443759} {\bibfield  {journal}
  {\bibinfo  {journal} {IEEE J. Sel. Areas Commun.}\ }\textbf {\bibinfo
  {volume} {42}},\ \bibinfo {pages} {2992} (\bibinfo {year}
  {2024})}\BibitemShut {NoStop}%
\bibitem [{\citenamefont {Louren{\c{c}}o}\ \emph {et~al.}(2019)\citenamefont
  {Louren{\c{c}}o}, \citenamefont {Martin},\ and\ \citenamefont
  {St{\"u}tzle}}]{LoMaSt19}%
  \BibitemOpen
  \bibfield  {author} {\bibinfo {author} {\bibfnamefont {H.~R.}\ \bibnamefont
  {Louren{\c{c}}o}}, \bibinfo {author} {\bibfnamefont {O.~C.}\ \bibnamefont
  {Martin}},\ and\ \bibinfo {author} {\bibfnamefont {T.}~\bibnamefont
  {St{\"u}tzle}},\ }\bibinfo {title} {{Iterated Local Search: Framework and
  Applications}},\ in\ \href {https://doi.org/10.1007/978-3-319-91086-4_5}
  {\emph {\bibinfo {booktitle} {Handbook of Metaheuristics}}},\ \bibinfo
  {editor} {edited by\ \bibinfo {editor} {\bibfnamefont {M.}~\bibnamefont
  {Gendreau}}\ and\ \bibinfo {editor} {\bibfnamefont {J.-Y.}\ \bibnamefont
  {Potvin}}}\ (\bibinfo  {publisher} {Springer International Publishing},\
  \bibinfo {address} {Cham},\ \bibinfo {year} {2019})\ pp.\ \bibinfo {pages}
  {129--168}\BibitemShut {NoStop}%
\bibitem [{\citenamefont {Kadowaki}\ and\ \citenamefont
  {Nishimori}(1998)}]{KaNi98}%
  \BibitemOpen
  \bibfield  {author} {\bibinfo {author} {\bibfnamefont {T.}~\bibnamefont
  {Kadowaki}}\ and\ \bibinfo {author} {\bibfnamefont {H.}~\bibnamefont
  {Nishimori}},\ }\bibfield  {title} {\bibinfo {title} {{Quantum annealing in
  the transverse Ising model}},\ }\href
  {https://doi.org/10.1103/PhysRevE.58.5355} {\bibfield  {journal} {\bibinfo
  {journal} {Phys. Rev. E}\ }\textbf {\bibinfo {volume} {58}},\ \bibinfo
  {pages} {5355} (\bibinfo {year} {1998})}\BibitemShut {NoStop}%
\bibitem [{\citenamefont {Farhi}\ \emph {et~al.}(2000)\citenamefont {Farhi},
  \citenamefont {Goldstone}, \citenamefont {Gutmann},\ and\ \citenamefont
  {Sipser}}]{FaEtAl00}%
  \BibitemOpen
  \bibfield  {author} {\bibinfo {author} {\bibfnamefont {E.}~\bibnamefont
  {Farhi}}, \bibinfo {author} {\bibfnamefont {J.}~\bibnamefont {Goldstone}},
  \bibinfo {author} {\bibfnamefont {S.}~\bibnamefont {Gutmann}},\ and\ \bibinfo
  {author} {\bibfnamefont {M.}~\bibnamefont {Sipser}},\ }\href@noop {}
  {\bibinfo {title} {{Quantum Computation by Adiabatic Evolution}}} (\bibinfo
  {year} {2000}),\ \Eprint {https://arxiv.org/abs/quant-ph/0001106}
  {arXiv:quant-ph/0001106 [quant-ph]} \BibitemShut {NoStop}%
\bibitem [{\citenamefont {Das}\ and\ \citenamefont
  {Chakrabarti}(2008)}]{DaCh08}%
  \BibitemOpen
  \bibfield  {author} {\bibinfo {author} {\bibfnamefont {A.}~\bibnamefont
  {Das}}\ and\ \bibinfo {author} {\bibfnamefont {B.~K.}\ \bibnamefont
  {Chakrabarti}},\ }\bibfield  {title} {\bibinfo {title} {{Colloquium: Quantum
  annealing and analog quantum computation}},\ }\href
  {https://doi.org/10.1103/RevModPhys.80.1061} {\bibfield  {journal} {\bibinfo
  {journal} {Rev. Mod. Phys.}\ }\textbf {\bibinfo {volume} {80}},\ \bibinfo
  {pages} {1061} (\bibinfo {year} {2008})}\BibitemShut {NoStop}%
\bibitem [{\citenamefont {Albash}\ and\ \citenamefont {Lidar}(2018)}]{AlLi18}%
  \BibitemOpen
  \bibfield  {author} {\bibinfo {author} {\bibfnamefont {T.}~\bibnamefont
  {Albash}}\ and\ \bibinfo {author} {\bibfnamefont {D.~A.}\ \bibnamefont
  {Lidar}},\ }\bibfield  {title} {\bibinfo {title} {Adiabatic quantum
  computation},\ }\href {https://doi.org/10.1103/RevModPhys.90.015002}
  {\bibfield  {journal} {\bibinfo  {journal} {Rev. Mod. Phys.}\ }\textbf
  {\bibinfo {volume} {90}},\ \bibinfo {pages} {015002} (\bibinfo {year}
  {2018})}\BibitemShut {NoStop}%
\bibitem [{\citenamefont {Hauke}\ \emph {et~al.}(2020)\citenamefont {Hauke},
  \citenamefont {Katzgraber}, \citenamefont {Lechner}, \citenamefont
  {Nishimori},\ and\ \citenamefont {Oliver}}]{HaEtAl20}%
  \BibitemOpen
  \bibfield  {author} {\bibinfo {author} {\bibfnamefont {P.}~\bibnamefont
  {Hauke}}, \bibinfo {author} {\bibfnamefont {H.~G.}\ \bibnamefont
  {Katzgraber}}, \bibinfo {author} {\bibfnamefont {W.}~\bibnamefont {Lechner}},
  \bibinfo {author} {\bibfnamefont {H.}~\bibnamefont {Nishimori}},\ and\
  \bibinfo {author} {\bibfnamefont {W.~D.}\ \bibnamefont {Oliver}},\ }\bibfield
   {title} {\bibinfo {title} {Perspectives of quantum annealing: methods and
  implementations},\ }\href {https://doi.org/10.1088/1361-6633/ab85b8}
  {\bibfield  {journal} {\bibinfo  {journal} {Rep. Prog. Phys.}\ }\textbf
  {\bibinfo {volume} {83}},\ \bibinfo {pages} {054401} (\bibinfo {year}
  {2020})}\BibitemShut {NoStop}%
\bibitem [{\citenamefont {Cerezo}\ \emph {et~al.}(2021)\citenamefont {Cerezo},
  \citenamefont {Arrasmith}, \citenamefont {Babbush}, \citenamefont {Benjamin},
  \citenamefont {Endo}, \citenamefont {Fujii}, \citenamefont {McClean},
  \citenamefont {Mitarai}, \citenamefont {Yuan}, \citenamefont {Cincio},\ and\
  \citenamefont {Coles}}]{CeEtAl21}%
  \BibitemOpen
  \bibfield  {author} {\bibinfo {author} {\bibfnamefont {M.}~\bibnamefont
  {Cerezo}}, \bibinfo {author} {\bibfnamefont {A.}~\bibnamefont {Arrasmith}},
  \bibinfo {author} {\bibfnamefont {R.}~\bibnamefont {Babbush}}, \bibinfo
  {author} {\bibfnamefont {S.~C.}\ \bibnamefont {Benjamin}}, \bibinfo {author}
  {\bibfnamefont {S.}~\bibnamefont {Endo}}, \bibinfo {author} {\bibfnamefont
  {K.}~\bibnamefont {Fujii}}, \bibinfo {author} {\bibfnamefont {J.~R.}\
  \bibnamefont {McClean}}, \bibinfo {author} {\bibfnamefont {K.}~\bibnamefont
  {Mitarai}}, \bibinfo {author} {\bibfnamefont {X.}~\bibnamefont {Yuan}},
  \bibinfo {author} {\bibfnamefont {L.}~\bibnamefont {Cincio}},\ and\ \bibinfo
  {author} {\bibfnamefont {P.~J.}\ \bibnamefont {Coles}},\ }\bibfield  {title}
  {\bibinfo {title} {Variational quantum algorithms},\ }\href
  {https://doi.org/10.1038/s42254-021-00348-9} {\bibfield  {journal} {\bibinfo
  {journal} {Nat. Rev. Phys.}\ }\textbf {\bibinfo {volume} {3}},\ \bibinfo
  {pages} {625} (\bibinfo {year} {2021})}\BibitemShut {NoStop}%
\bibitem [{\citenamefont {Bharti}\ \emph {et~al.}(2022)\citenamefont {Bharti},
  \citenamefont {Cervera-Lierta}, \citenamefont {Kyaw}, \citenamefont {Haug},
  \citenamefont {Alperin-Lea}, \citenamefont {Anand}, \citenamefont {Degroote},
  \citenamefont {Heimonen}, \citenamefont {Kottmann}, \citenamefont {Menke},
  \citenamefont {Mok}, \citenamefont {Sim}, \citenamefont {Kwek},\ and\
  \citenamefont {Aspuru-Guzik}}]{BhEtAl22}%
  \BibitemOpen
  \bibfield  {author} {\bibinfo {author} {\bibfnamefont {K.}~\bibnamefont
  {Bharti}}, \bibinfo {author} {\bibfnamefont {A.}~\bibnamefont
  {Cervera-Lierta}}, \bibinfo {author} {\bibfnamefont {T.~H.}\ \bibnamefont
  {Kyaw}}, \bibinfo {author} {\bibfnamefont {T.}~\bibnamefont {Haug}}, \bibinfo
  {author} {\bibfnamefont {S.}~\bibnamefont {Alperin-Lea}}, \bibinfo {author}
  {\bibfnamefont {A.}~\bibnamefont {Anand}}, \bibinfo {author} {\bibfnamefont
  {M.}~\bibnamefont {Degroote}}, \bibinfo {author} {\bibfnamefont
  {H.}~\bibnamefont {Heimonen}}, \bibinfo {author} {\bibfnamefont {J.~S.}\
  \bibnamefont {Kottmann}}, \bibinfo {author} {\bibfnamefont {T.}~\bibnamefont
  {Menke}}, \bibinfo {author} {\bibfnamefont {W.-K.}\ \bibnamefont {Mok}},
  \bibinfo {author} {\bibfnamefont {S.}~\bibnamefont {Sim}}, \bibinfo {author}
  {\bibfnamefont {L.-C.}\ \bibnamefont {Kwek}},\ and\ \bibinfo {author}
  {\bibfnamefont {A.}~\bibnamefont {Aspuru-Guzik}},\ }\bibfield  {title}
  {\bibinfo {title} {Noisy intermediate-scale quantum algorithms},\ }\href
  {https://doi.org/10.1103/RevModPhys.94.015004} {\bibfield  {journal}
  {\bibinfo  {journal} {Rev. Mod. Phys.}\ }\textbf {\bibinfo {volume} {94}},\
  \bibinfo {pages} {015004} (\bibinfo {year} {2022})}\BibitemShut {NoStop}%
\bibitem [{\citenamefont {Tilly}\ \emph {et~al.}(2022)\citenamefont {Tilly},
  \citenamefont {Chen}, \citenamefont {Cao}, \citenamefont {Picozzi},
  \citenamefont {Setia}, \citenamefont {Li}, \citenamefont {Grant},
  \citenamefont {Wossnig}, \citenamefont {Rungger}, \citenamefont {Booth},\
  and\ \citenamefont {Tennyson}}]{TiEtAl22}%
  \BibitemOpen
  \bibfield  {author} {\bibinfo {author} {\bibfnamefont {J.}~\bibnamefont
  {Tilly}}, \bibinfo {author} {\bibfnamefont {H.}~\bibnamefont {Chen}},
  \bibinfo {author} {\bibfnamefont {S.}~\bibnamefont {Cao}}, \bibinfo {author}
  {\bibfnamefont {D.}~\bibnamefont {Picozzi}}, \bibinfo {author} {\bibfnamefont
  {K.}~\bibnamefont {Setia}}, \bibinfo {author} {\bibfnamefont
  {Y.}~\bibnamefont {Li}}, \bibinfo {author} {\bibfnamefont {E.}~\bibnamefont
  {Grant}}, \bibinfo {author} {\bibfnamefont {L.}~\bibnamefont {Wossnig}},
  \bibinfo {author} {\bibfnamefont {I.}~\bibnamefont {Rungger}}, \bibinfo
  {author} {\bibfnamefont {G.~H.}\ \bibnamefont {Booth}},\ and\ \bibinfo
  {author} {\bibfnamefont {J.}~\bibnamefont {Tennyson}},\ }\bibfield  {title}
  {\bibinfo {title} {{The Variational Quantum Eigensolver: A review of methods
  and best practices}},\ }\href
  {https://doi.org/https://doi.org/10.1016/j.physrep.2022.08.003} {\bibfield
  {journal} {\bibinfo  {journal} {Phys. Rep.}\ }\textbf {\bibinfo {volume}
  {986}},\ \bibinfo {pages} {1} (\bibinfo {year} {2022})}\BibitemShut {NoStop}%
\bibitem [{\citenamefont {Farhi}\ \emph {et~al.}(2014)\citenamefont {Farhi},
  \citenamefont {Goldstone},\ and\ \citenamefont {Gutmann}}]{FaGoGu14}%
  \BibitemOpen
  \bibfield  {author} {\bibinfo {author} {\bibfnamefont {E.}~\bibnamefont
  {Farhi}}, \bibinfo {author} {\bibfnamefont {J.}~\bibnamefont {Goldstone}},\
  and\ \bibinfo {author} {\bibfnamefont {S.}~\bibnamefont {Gutmann}},\
  }\href@noop {} {\bibinfo {title} {{A Quantum Approximate Optimization
  Algorithm}}} (\bibinfo {year} {2014}),\ \Eprint
  {https://arxiv.org/abs/1411.4028} {arXiv:1411.4028 [quant-ph]} \BibitemShut
  {NoStop}%
\bibitem [{\citenamefont {Zhou}\ \emph
  {et~al.}(2020{\natexlab{a}})\citenamefont {Zhou}, \citenamefont {Wang},
  \citenamefont {Choi}, \citenamefont {Pichler},\ and\ \citenamefont
  {Lukin}}]{ZhEtAl20}%
  \BibitemOpen
  \bibfield  {author} {\bibinfo {author} {\bibfnamefont {L.}~\bibnamefont
  {Zhou}}, \bibinfo {author} {\bibfnamefont {S.-T.}\ \bibnamefont {Wang}},
  \bibinfo {author} {\bibfnamefont {S.}~\bibnamefont {Choi}}, \bibinfo {author}
  {\bibfnamefont {H.}~\bibnamefont {Pichler}},\ and\ \bibinfo {author}
  {\bibfnamefont {M.~D.}\ \bibnamefont {Lukin}},\ }\bibfield  {title} {\bibinfo
  {title} {{Quantum Approximate Optimization Algorithm: Performance, Mechanism,
  and Implementation on Near-Term Devices}},\ }\href
  {https://doi.org/10.1103/PhysRevX.10.021067} {\bibfield  {journal} {\bibinfo
  {journal} {Phys. Rev. X}\ }\textbf {\bibinfo {volume} {10}},\ \bibinfo
  {pages} {021067} (\bibinfo {year} {2020}{\natexlab{a}})}\BibitemShut
  {NoStop}%
\bibitem [{\citenamefont {Perez-Garcia}\ \emph {et~al.}(2007)\citenamefont
  {Perez-Garcia}, \citenamefont {Verstraete}, \citenamefont {Wolf},\ and\
  \citenamefont {Cirac}}]{PeEtAl07}%
  \BibitemOpen
  \bibfield  {author} {\bibinfo {author} {\bibfnamefont {D.}~\bibnamefont
  {Perez-Garcia}}, \bibinfo {author} {\bibfnamefont {F.}~\bibnamefont
  {Verstraete}}, \bibinfo {author} {\bibfnamefont {M.~M.}\ \bibnamefont
  {Wolf}},\ and\ \bibinfo {author} {\bibfnamefont {J.~I.}\ \bibnamefont
  {Cirac}},\ }\bibfield  {title} {\bibinfo {title} {Matrix product state
  representations},\ }\href {https://doi.org/10.26421/QIC7.5-6-1} {\bibfield
  {journal} {\bibinfo  {journal} {Quantum Inf. Comput.}\ }\textbf {\bibinfo
  {volume} {7}},\ \bibinfo {pages} {401} (\bibinfo {year} {2007})}\BibitemShut
  {NoStop}%
\bibitem [{\citenamefont {Verstraete}\ \emph {et~al.}(2008)\citenamefont
  {Verstraete}, \citenamefont {Murg},\ and\ \citenamefont {Cirac}}]{VeMuCi08}%
  \BibitemOpen
  \bibfield  {author} {\bibinfo {author} {\bibfnamefont {F.}~\bibnamefont
  {Verstraete}}, \bibinfo {author} {\bibfnamefont {V.}~\bibnamefont {Murg}},\
  and\ \bibinfo {author} {\bibfnamefont {J.~I.}\ \bibnamefont {Cirac}},\
  }\bibfield  {title} {\bibinfo {title} {Matrix product states, projected
  entangled pair states, and variational renormalization group methods for
  quantum spin systems},\ }\href {https://doi.org/10.1080/14789940801912366}
  {\bibfield  {journal} {\bibinfo  {journal} {Adv. Phys.}\ }\textbf {\bibinfo
  {volume} {57}},\ \bibinfo {pages} {143} (\bibinfo {year} {2008})}\BibitemShut
  {NoStop}%
\bibitem [{\citenamefont {Schollw{\"o}ck}(2011)}]{Sc11}%
  \BibitemOpen
  \bibfield  {author} {\bibinfo {author} {\bibfnamefont {U.}~\bibnamefont
  {Schollw{\"o}ck}},\ }\bibfield  {title} {\bibinfo {title} {The density-matrix
  renormalization group in the age of matrix product states},\ }\href
  {https://doi.org/https://doi.org/10.1016/j.aop.2010.09.012} {\bibfield
  {journal} {\bibinfo  {journal} {Ann. Phys. (N. Y.)}\ }\textbf {\bibinfo
  {volume} {326}},\ \bibinfo {pages} {96} (\bibinfo {year} {2011})},\ \bibinfo
  {note} {january 2011 Special Issue}\BibitemShut {NoStop}%
\bibitem [{\citenamefont {Or\'{u}s}(2014)}]{Or14}%
  \BibitemOpen
  \bibfield  {author} {\bibinfo {author} {\bibfnamefont {R.}~\bibnamefont
  {Or\'{u}s}},\ }\bibfield  {title} {\bibinfo {title} {{A practical
  introduction to tensor networks: Matrix product states and projected
  entangled pair states}},\ }\href
  {https://doi.org/https://doi.org/10.1016/j.aop.2014.06.013} {\bibfield
  {journal} {\bibinfo  {journal} {Ann. Phys. (N. Y.)}\ }\textbf {\bibinfo
  {volume} {349}},\ \bibinfo {pages} {117} (\bibinfo {year}
  {2014})}\BibitemShut {NoStop}%
\bibitem [{\citenamefont {Ba\~{n}uls}(2023)}]{Ba23}%
  \BibitemOpen
  \bibfield  {author} {\bibinfo {author} {\bibfnamefont {M.~C.}\ \bibnamefont
  {Ba\~{n}uls}},\ }\bibfield  {title} {\bibinfo {title} {{Tensor Network
  Algorithms: A Route Map}},\ }\href
  {https://doi.org/10.1146/annurev-conmatphys-040721-022705} {\bibfield
  {journal} {\bibinfo  {journal} {Annu. Rev. Condens. Matter Phys.}\ }\textbf
  {\bibinfo {volume} {14}},\ \bibinfo {pages} {173} (\bibinfo {year}
  {2023})}\BibitemShut {NoStop}%
\bibitem [{\citenamefont {Zhou}\ \emph
  {et~al.}(2020{\natexlab{b}})\citenamefont {Zhou}, \citenamefont
  {Stoudenmire},\ and\ \citenamefont {Waintal}}]{ZhStWa20}%
  \BibitemOpen
  \bibfield  {author} {\bibinfo {author} {\bibfnamefont {Y.}~\bibnamefont
  {Zhou}}, \bibinfo {author} {\bibfnamefont {E.~M.}\ \bibnamefont
  {Stoudenmire}},\ and\ \bibinfo {author} {\bibfnamefont {X.}~\bibnamefont
  {Waintal}},\ }\bibfield  {title} {\bibinfo {title} {{What Limits the
  Simulation of Quantum Computers?}},\ }\href
  {https://doi.org/10.1103/PhysRevX.10.041038} {\bibfield  {journal} {\bibinfo
  {journal} {Phys. Rev. X}\ }\textbf {\bibinfo {volume} {10}},\ \bibinfo
  {pages} {041038} (\bibinfo {year} {2020}{\natexlab{b}})}\BibitemShut
  {NoStop}%
\bibitem [{\citenamefont {Ayral}\ \emph {et~al.}(2023)\citenamefont {Ayral},
  \citenamefont {Louvet}, \citenamefont {Zhou}, \citenamefont {Lambert},
  \citenamefont {Stoudenmire},\ and\ \citenamefont {Waintal}}]{AyEtAl23}%
  \BibitemOpen
  \bibfield  {author} {\bibinfo {author} {\bibfnamefont {T.}~\bibnamefont
  {Ayral}}, \bibinfo {author} {\bibfnamefont {T.}~\bibnamefont {Louvet}},
  \bibinfo {author} {\bibfnamefont {Y.}~\bibnamefont {Zhou}}, \bibinfo {author}
  {\bibfnamefont {C.}~\bibnamefont {Lambert}}, \bibinfo {author} {\bibfnamefont
  {E.~M.}\ \bibnamefont {Stoudenmire}},\ and\ \bibinfo {author} {\bibfnamefont
  {X.}~\bibnamefont {Waintal}},\ }\bibfield  {title} {\bibinfo {title}
  {{Density-Matrix Renormalization Group Algorithm for Simulating Quantum
  Circuits with a Finite Fidelity}},\ }\href
  {https://doi.org/10.1103/PRXQuantum.4.020304} {\bibfield  {journal} {\bibinfo
   {journal} {PRX Quantum}\ }\textbf {\bibinfo {volume} {4}},\ \bibinfo {pages}
  {020304} (\bibinfo {year} {2023})}\BibitemShut {NoStop}%
\bibitem [{\citenamefont {Karp}(1972)}]{Ka72}%
  \BibitemOpen
  \bibfield  {author} {\bibinfo {author} {\bibfnamefont {R.~M.}\ \bibnamefont
  {Karp}},\ }\bibinfo {title} {{Reducibility among Combinatorial Problems}},\
  in\ \href {https://doi.org/10.1007/978-1-4684-2001-2_9} {\emph {\bibinfo
  {booktitle} {Complexity of Computer Computations. The IBM Research Symposia
  Series.}}},\ \bibinfo {editor} {edited by\ \bibinfo {editor} {\bibfnamefont
  {R.~E.}\ \bibnamefont {Miller}}, \bibinfo {editor} {\bibfnamefont {J.~W.}\
  \bibnamefont {Thatcher}},\ and\ \bibinfo {editor} {\bibfnamefont {J.~D.}\
  \bibnamefont {Bohlinger}}}\ (\bibinfo  {publisher} {Springer US},\ \bibinfo
  {address} {Boston, MA},\ \bibinfo {year} {1972})\ pp.\ \bibinfo {pages}
  {85--103}\BibitemShut {NoStop}%
\bibitem [{\citenamefont {Goemans}\ and\ \citenamefont
  {Williamson}(1995)}]{GoWi95}%
  \BibitemOpen
  \bibfield  {author} {\bibinfo {author} {\bibfnamefont {M.~X.}\ \bibnamefont
  {Goemans}}\ and\ \bibinfo {author} {\bibfnamefont {D.~P.}\ \bibnamefont
  {Williamson}},\ }\bibfield  {title} {\bibinfo {title} {Improved approximation
  algorithms for maximum cut and satisfiability problems using semidefinite
  programming},\ }\href {https://doi.org/10.1145/227683.227684} {\bibfield
  {journal} {\bibinfo  {journal} {J. ACM}\ }\textbf {\bibinfo {volume} {42}},\
  \bibinfo {pages} {1115} (\bibinfo {year} {1995})}\BibitemShut {NoStop}%
\bibitem [{\citenamefont {Bowles}\ \emph {et~al.}(2022)\citenamefont {Bowles},
  \citenamefont {Dauphin}, \citenamefont {Huembeli}, \citenamefont {Martinez},\
  and\ \citenamefont {Ac\'{\i}n}}]{BoEtAl22}%
  \BibitemOpen
  \bibfield  {author} {\bibinfo {author} {\bibfnamefont {J.}~\bibnamefont
  {Bowles}}, \bibinfo {author} {\bibfnamefont {A.}~\bibnamefont {Dauphin}},
  \bibinfo {author} {\bibfnamefont {P.}~\bibnamefont {Huembeli}}, \bibinfo
  {author} {\bibfnamefont {J.}~\bibnamefont {Martinez}},\ and\ \bibinfo
  {author} {\bibfnamefont {A.}~\bibnamefont {Ac\'{\i}n}},\ }\bibfield  {title}
  {\bibinfo {title} {{Quadratic Unconstrained Binary Optimization via
  Quantum-Inspired Annealing}},\ }\href
  {https://doi.org/10.1103/PhysRevApplied.18.034016} {\bibfield  {journal}
  {\bibinfo  {journal} {Phys. Rev. Appl.}\ }\textbf {\bibinfo {volume} {18}},\
  \bibinfo {pages} {034016} (\bibinfo {year} {2022})}\BibitemShut {NoStop}%
\bibitem [{\citenamefont {Guaita}\ \emph {et~al.}(2021)\citenamefont {Guaita},
  \citenamefont {Hackl}, \citenamefont {Shi}, \citenamefont {Demler},\ and\
  \citenamefont {Cirac}}]{GuEtAl21}%
  \BibitemOpen
  \bibfield  {author} {\bibinfo {author} {\bibfnamefont {T.}~\bibnamefont
  {Guaita}}, \bibinfo {author} {\bibfnamefont {L.}~\bibnamefont {Hackl}},
  \bibinfo {author} {\bibfnamefont {T.}~\bibnamefont {Shi}}, \bibinfo {author}
  {\bibfnamefont {E.}~\bibnamefont {Demler}},\ and\ \bibinfo {author}
  {\bibfnamefont {J.~I.}\ \bibnamefont {Cirac}},\ }\bibfield  {title} {\bibinfo
  {title} {Generalization of group-theoretic coherent states for variational
  calculations},\ }\href {https://doi.org/10.1103/PhysRevResearch.3.023090}
  {\bibfield  {journal} {\bibinfo  {journal} {Phys. Rev. Res.}\ }\textbf
  {\bibinfo {volume} {3}},\ \bibinfo {pages} {023090} (\bibinfo {year}
  {2021})}\BibitemShut {NoStop}%
\bibitem [{\citenamefont {Schindler}\ \emph {et~al.}(2022)\citenamefont
  {Schindler}, \citenamefont {Guaita}, \citenamefont {Shi}, \citenamefont
  {Demler},\ and\ \citenamefont {Cirac}}]{ScEtAl22}%
  \BibitemOpen
  \bibfield  {author} {\bibinfo {author} {\bibfnamefont {P.~M.}\ \bibnamefont
  {Schindler}}, \bibinfo {author} {\bibfnamefont {T.}~\bibnamefont {Guaita}},
  \bibinfo {author} {\bibfnamefont {T.}~\bibnamefont {Shi}}, \bibinfo {author}
  {\bibfnamefont {E.}~\bibnamefont {Demler}},\ and\ \bibinfo {author}
  {\bibfnamefont {J.~I.}\ \bibnamefont {Cirac}},\ }\bibfield  {title} {\bibinfo
  {title} {{Variational Ansatz for the Ground State of the Quantum
  Sherrington-Kirkpatrick Model}},\ }\href
  {https://doi.org/10.1103/PhysRevLett.129.220401} {\bibfield  {journal}
  {\bibinfo  {journal} {Phys. Rev. Lett.}\ }\textbf {\bibinfo {volume} {129}},\
  \bibinfo {pages} {220401} (\bibinfo {year} {2022})}\BibitemShut {NoStop}%
\bibitem [{\citenamefont {Fioroni}\ and\ \citenamefont
  {Savona}(2025)}]{FiSa25}%
  \BibitemOpen
  \bibfield  {author} {\bibinfo {author} {\bibfnamefont {L.}~\bibnamefont
  {Fioroni}}\ and\ \bibinfo {author} {\bibfnamefont {V.}~\bibnamefont
  {Savona}},\ }\bibfield  {title} {\bibinfo {title} {Entanglement-assisted
  variational algorithm for discrete optimization problems},\ }\href
  {https://doi.org/10.1038/s42005-025-02338-0} {\bibfield  {journal} {\bibinfo
  {journal} {Commun. Phys.}\ }\textbf {\bibinfo {volume} {8}},\ \bibinfo
  {pages} {438} (\bibinfo {year} {2025})}\BibitemShut {NoStop}%
\bibitem [{\citenamefont {Ye}(2003)}]{Gset}%
  \BibitemOpen
  \bibfield  {author} {\bibinfo {author} {\bibfnamefont {Y.}~\bibnamefont
  {Ye}},\ }\href@noop {} {\bibinfo {title} {Gset dataset}},\ \bibinfo
  {howpublished} {\url{https://web.stanford.edu/~yyye/yyye/Gset/}} (\bibinfo
  {year} {2003})\BibitemShut {NoStop}%
\bibitem [{\citenamefont {Ba\~nuls}\ \emph {et~al.}(2006)\citenamefont
  {Ba\~nuls}, \citenamefont {Or\'us}, \citenamefont {Latorre}, \citenamefont
  {P\'erez},\ and\ \citenamefont {Ruiz-Femen\'{\i}a}}]{BaEtAl06}%
  \BibitemOpen
  \bibfield  {author} {\bibinfo {author} {\bibfnamefont {M.~C.}\ \bibnamefont
  {Ba\~nuls}}, \bibinfo {author} {\bibfnamefont {R.}~\bibnamefont {Or\'us}},
  \bibinfo {author} {\bibfnamefont {J.~I.}\ \bibnamefont {Latorre}}, \bibinfo
  {author} {\bibfnamefont {A.}~\bibnamefont {P\'erez}},\ and\ \bibinfo {author}
  {\bibfnamefont {P.}~\bibnamefont {Ruiz-Femen\'{\i}a}},\ }\bibfield  {title}
  {\bibinfo {title} {Simulation of many-qubit quantum computation with matrix
  product states},\ }\href {https://doi.org/10.1103/PhysRevA.73.022344}
  {\bibfield  {journal} {\bibinfo  {journal} {Phys. Rev. A}\ }\textbf {\bibinfo
  {volume} {73}},\ \bibinfo {pages} {022344} (\bibinfo {year}
  {2006})}\BibitemShut {NoStop}%
\bibitem [{\citenamefont {Smolin}\ and\ \citenamefont {Smith}(2014)}]{SmSm14}%
  \BibitemOpen
  \bibfield  {author} {\bibinfo {author} {\bibfnamefont {J.~A.}\ \bibnamefont
  {Smolin}}\ and\ \bibinfo {author} {\bibfnamefont {G.}~\bibnamefont {Smith}},\
  }\bibfield  {title} {\bibinfo {title} {Classical signature of quantum
  annealing},\ }\href {https://doi.org/10.3389/fphy.2014.00052} {\bibfield
  {journal} {\bibinfo  {journal} {Front. Phys.}\ }\textbf {\bibinfo {volume}
  {2}},\ \bibinfo {pages} {52} (\bibinfo {year} {2014})}\BibitemShut {NoStop}%
\bibitem [{\citenamefont {Bauer}\ \emph {et~al.}(2015)\citenamefont {Bauer},
  \citenamefont {Wang}, \citenamefont {Pižorn},\ and\ \citenamefont
  {Troyer}}]{BaEtAl15}%
  \BibitemOpen
  \bibfield  {author} {\bibinfo {author} {\bibfnamefont {B.}~\bibnamefont
  {Bauer}}, \bibinfo {author} {\bibfnamefont {L.}~\bibnamefont {Wang}},
  \bibinfo {author} {\bibfnamefont {I.}~\bibnamefont {Pižorn}},\ and\ \bibinfo
  {author} {\bibfnamefont {M.}~\bibnamefont {Troyer}},\ }\href@noop {}
  {\bibinfo {title} {Entanglement as a resource in adiabatic quantum
  optimization}} (\bibinfo {year} {2015}),\ \Eprint
  {https://arxiv.org/abs/1501.06914} {arXiv:1501.06914 [cond-mat.dis-nn]}
  \BibitemShut {NoStop}%
\bibitem [{\citenamefont {Hatomura}\ and\ \citenamefont {Mori}(2018)}]{HaMo18}%
  \BibitemOpen
  \bibfield  {author} {\bibinfo {author} {\bibfnamefont {T.}~\bibnamefont
  {Hatomura}}\ and\ \bibinfo {author} {\bibfnamefont {T.}~\bibnamefont
  {Mori}},\ }\bibfield  {title} {\bibinfo {title} {Shortcuts to adiabatic
  classical spin dynamics mimicking quantum annealing},\ }\href
  {https://doi.org/10.1103/PhysRevE.98.032136} {\bibfield  {journal} {\bibinfo
  {journal} {Phys. Rev. E}\ }\textbf {\bibinfo {volume} {98}},\ \bibinfo
  {pages} {032136} (\bibinfo {year} {2018})}\BibitemShut {NoStop}%
\bibitem [{\citenamefont {Mugel}\ \emph {et~al.}(2022)\citenamefont {Mugel},
  \citenamefont {Kuchkovsky}, \citenamefont {S\'anchez}, \citenamefont
  {Fern\'andez-Lorenzo}, \citenamefont {Luis-Hita}, \citenamefont {Lizaso},\
  and\ \citenamefont {Or\'us}}]{MuEtAl22}%
  \BibitemOpen
  \bibfield  {author} {\bibinfo {author} {\bibfnamefont {S.}~\bibnamefont
  {Mugel}}, \bibinfo {author} {\bibfnamefont {C.}~\bibnamefont {Kuchkovsky}},
  \bibinfo {author} {\bibfnamefont {E.}~\bibnamefont {S\'anchez}}, \bibinfo
  {author} {\bibfnamefont {S.}~\bibnamefont {Fern\'andez-Lorenzo}}, \bibinfo
  {author} {\bibfnamefont {J.}~\bibnamefont {Luis-Hita}}, \bibinfo {author}
  {\bibfnamefont {E.}~\bibnamefont {Lizaso}},\ and\ \bibinfo {author}
  {\bibfnamefont {R.}~\bibnamefont {Or\'us}},\ }\bibfield  {title} {\bibinfo
  {title} {Dynamic portfolio optimization with real datasets using quantum
  processors and quantum-inspired tensor networks},\ }\href
  {https://doi.org/10.1103/PhysRevResearch.4.013006} {\bibfield  {journal}
  {\bibinfo  {journal} {Phys. Rev. Res.}\ }\textbf {\bibinfo {volume} {4}},\
  \bibinfo {pages} {013006} (\bibinfo {year} {2022})}\BibitemShut {NoStop}%
\bibitem [{\citenamefont {Sreedhar}\ \emph {et~al.}(2022)\citenamefont
  {Sreedhar}, \citenamefont {Vikstål}, \citenamefont {Svensson}, \citenamefont
  {Ask}, \citenamefont {Johansson},\ and\ \citenamefont
  {García-Álvarez}}]{SrEtAl22}%
  \BibitemOpen
  \bibfield  {author} {\bibinfo {author} {\bibfnamefont {R.}~\bibnamefont
  {Sreedhar}}, \bibinfo {author} {\bibfnamefont {P.}~\bibnamefont {Vikstål}},
  \bibinfo {author} {\bibfnamefont {M.}~\bibnamefont {Svensson}}, \bibinfo
  {author} {\bibfnamefont {A.}~\bibnamefont {Ask}}, \bibinfo {author}
  {\bibfnamefont {G.}~\bibnamefont {Johansson}},\ and\ \bibinfo {author}
  {\bibfnamefont {L.}~\bibnamefont {García-Álvarez}},\ }\href@noop {}
  {\bibinfo {title} {{The Quantum Approximate Optimization Algorithm
  performance with low entanglement and high circuit depth}}} (\bibinfo {year}
  {2022}),\ \Eprint {https://arxiv.org/abs/2207.03404} {arXiv:2207.03404
  [quant-ph]} \BibitemShut {NoStop}%
\bibitem [{\citenamefont {Dupont}\ \emph
  {et~al.}(2022{\natexlab{a}})\citenamefont {Dupont}, \citenamefont {Didier},
  \citenamefont {Hodson}, \citenamefont {Moore},\ and\ \citenamefont
  {Reagor}}]{DuEtAl22a}%
  \BibitemOpen
  \bibfield  {author} {\bibinfo {author} {\bibfnamefont {M.}~\bibnamefont
  {Dupont}}, \bibinfo {author} {\bibfnamefont {N.}~\bibnamefont {Didier}},
  \bibinfo {author} {\bibfnamefont {M.~J.}\ \bibnamefont {Hodson}}, \bibinfo
  {author} {\bibfnamefont {J.~E.}\ \bibnamefont {Moore}},\ and\ \bibinfo
  {author} {\bibfnamefont {M.~J.}\ \bibnamefont {Reagor}},\ }\bibfield  {title}
  {\bibinfo {title} {Entanglement perspective on the quantum approximate
  optimization algorithm},\ }\href
  {https://doi.org/10.1103/PhysRevA.106.022423} {\bibfield  {journal} {\bibinfo
   {journal} {Phys. Rev. A}\ }\textbf {\bibinfo {volume} {106}},\ \bibinfo
  {pages} {022423} (\bibinfo {year} {2022}{\natexlab{a}})}\BibitemShut
  {NoStop}%
\bibitem [{\citenamefont {Veszeli}\ and\ \citenamefont
  {Vattay}(2022)}]{VeVa22}%
  \BibitemOpen
  \bibfield  {author} {\bibinfo {author} {\bibfnamefont {M.~T.}\ \bibnamefont
  {Veszeli}}\ and\ \bibinfo {author} {\bibfnamefont {G.}~\bibnamefont
  {Vattay}},\ }\bibfield  {title} {\bibinfo {title} {{Mean field approximation
  for solving QUBO problems}},\ }\href
  {https://doi.org/10.1371/journal.pone.0273709} {\bibfield  {journal}
  {\bibinfo  {journal} {PLOS ONE}\ }\textbf {\bibinfo {volume} {17}},\ \bibinfo
  {pages} {1} (\bibinfo {year} {2022})}\BibitemShut {NoStop}%
\bibitem [{\citenamefont {Dupont}\ \emph
  {et~al.}(2022{\natexlab{b}})\citenamefont {Dupont}, \citenamefont {Didier},
  \citenamefont {Hodson}, \citenamefont {Moore},\ and\ \citenamefont
  {Reagor}}]{DuEtAl22b}%
  \BibitemOpen
  \bibfield  {author} {\bibinfo {author} {\bibfnamefont {M.}~\bibnamefont
  {Dupont}}, \bibinfo {author} {\bibfnamefont {N.}~\bibnamefont {Didier}},
  \bibinfo {author} {\bibfnamefont {M.~J.}\ \bibnamefont {Hodson}}, \bibinfo
  {author} {\bibfnamefont {J.~E.}\ \bibnamefont {Moore}},\ and\ \bibinfo
  {author} {\bibfnamefont {M.~J.}\ \bibnamefont {Reagor}},\ }\bibfield  {title}
  {\bibinfo {title} {{Calibrating the Classical Hardness of the Quantum
  Approximate Optimization Algorithm}},\ }\href
  {https://doi.org/10.1103/PRXQuantum.3.040339} {\bibfield  {journal} {\bibinfo
   {journal} {PRX Quantum}\ }\textbf {\bibinfo {volume} {3}},\ \bibinfo {pages}
  {040339} (\bibinfo {year} {2022}{\natexlab{b}})}\BibitemShut {NoStop}%
\bibitem [{\citenamefont {Lami}\ \emph {et~al.}(2023)\citenamefont {Lami},
  \citenamefont {Torta}, \citenamefont {Santoro},\ and\ \citenamefont
  {Collura}}]{LaEtAl23}%
  \BibitemOpen
  \bibfield  {author} {\bibinfo {author} {\bibfnamefont {G.}~\bibnamefont
  {Lami}}, \bibinfo {author} {\bibfnamefont {P.}~\bibnamefont {Torta}},
  \bibinfo {author} {\bibfnamefont {G.~E.}\ \bibnamefont {Santoro}},\ and\
  \bibinfo {author} {\bibfnamefont {M.}~\bibnamefont {Collura}},\ }\bibfield
  {title} {\bibinfo {title} {{Quantum annealing for neural network optimization
  problems: A new approach via tensor network simulations}},\ }\href
  {https://doi.org/10.21468/SciPostPhys.14.5.117} {\bibfield  {journal}
  {\bibinfo  {journal} {SciPost Phys.}\ }\textbf {\bibinfo {volume} {14}},\
  \bibinfo {pages} {117} (\bibinfo {year} {2023})}\BibitemShut {NoStop}%
\bibitem [{\citenamefont {Lopez-Piqueres}\ \emph {et~al.}(2023)\citenamefont
  {Lopez-Piqueres}, \citenamefont {Chen},\ and\ \citenamefont
  {Perdomo-Ortiz}}]{LoChPe23}%
  \BibitemOpen
  \bibfield  {author} {\bibinfo {author} {\bibfnamefont {J.}~\bibnamefont
  {Lopez-Piqueres}}, \bibinfo {author} {\bibfnamefont {J.}~\bibnamefont
  {Chen}},\ and\ \bibinfo {author} {\bibfnamefont {A.}~\bibnamefont
  {Perdomo-Ortiz}},\ }\bibfield  {title} {\bibinfo {title} {Symmetric tensor
  networks for generative modeling and constrained combinatorial
  optimization},\ }\href {https://doi.org/10.1088/2632-2153/ace0f5} {\bibfield
  {journal} {\bibinfo  {journal} {Mach. Learn.: Sci. Technol.}\ }\textbf
  {\bibinfo {volume} {4}},\ \bibinfo {pages} {035009} (\bibinfo {year}
  {2023})}\BibitemShut {NoStop}%
\bibitem [{\citenamefont {Batsheva}\ \emph {et~al.}(2023)\citenamefont
  {Batsheva}, \citenamefont {Chertkov}, \citenamefont {Ryzhakov},\ and\
  \citenamefont {Oseledets}}]{BaEtAl23}%
  \BibitemOpen
  \bibfield  {author} {\bibinfo {author} {\bibfnamefont {A.}~\bibnamefont
  {Batsheva}}, \bibinfo {author} {\bibfnamefont {A.}~\bibnamefont {Chertkov}},
  \bibinfo {author} {\bibfnamefont {G.}~\bibnamefont {Ryzhakov}},\ and\
  \bibinfo {author} {\bibfnamefont {I.}~\bibnamefont {Oseledets}},\ }\bibfield
  {title} {\bibinfo {title} {{PROTES: Probabilistic Optimization with Tensor
  Sampling}},\ }in\ \href {https://openreview.net/forum?id=R9R7YDOar1} {\emph
  {\bibinfo {booktitle} {37th Conference on Neural Information Processing
  Systems (NeurIPS)}}}\ (\bibinfo  {publisher} {Curran Associates, Inc.},\
  \bibinfo {address} {Red Hook, New York},\ \bibinfo {year} {2023})\BibitemShut
  {NoStop}%
\bibitem [{\citenamefont {Alcazar}\ \emph {et~al.}(2024)\citenamefont
  {Alcazar}, \citenamefont {Ghazi~Vakili}, \citenamefont {Kalayci},\ and\
  \citenamefont {Perdomo-Ortiz}}]{AlEtAl24}%
  \BibitemOpen
  \bibfield  {author} {\bibinfo {author} {\bibfnamefont {J.}~\bibnamefont
  {Alcazar}}, \bibinfo {author} {\bibfnamefont {M.}~\bibnamefont
  {Ghazi~Vakili}}, \bibinfo {author} {\bibfnamefont {C.~B.}\ \bibnamefont
  {Kalayci}},\ and\ \bibinfo {author} {\bibfnamefont {A.}~\bibnamefont
  {Perdomo-Ortiz}},\ }\bibfield  {title} {\bibinfo {title} {Enhancing
  combinatorial optimization with classical and quantum generative models},\
  }\href {https://doi.org/10.1038/s41467-024-46959-5} {\bibfield  {journal}
  {\bibinfo  {journal} {Nat. Commun.}\ }\textbf {\bibinfo {volume} {15}},\
  \bibinfo {pages} {2761} (\bibinfo {year} {2024})}\BibitemShut {NoStop}%
\bibitem [{\citenamefont {Gardiner}\ and\ \citenamefont
  {Lopez-Piqueres}(2024)}]{GaLo24}%
  \BibitemOpen
  \bibfield  {author} {\bibinfo {author} {\bibfnamefont {J.}~\bibnamefont
  {Gardiner}}\ and\ \bibinfo {author} {\bibfnamefont {J.}~\bibnamefont
  {Lopez-Piqueres}},\ }\href@noop {} {\bibinfo {title} {{Tensor Network
  Estimation of Distribution Algorithms}}} (\bibinfo {year} {2024}),\ \Eprint
  {https://arxiv.org/abs/2412.19780} {arXiv:2412.19780 [cs.LG]} \BibitemShut
  {NoStop}%
\bibitem [{\citenamefont {Lopez-Piqueres}\ and\ \citenamefont
  {Chen}(2025)}]{LoCh25}%
  \BibitemOpen
  \bibfield  {author} {\bibinfo {author} {\bibfnamefont {J.}~\bibnamefont
  {Lopez-Piqueres}}\ and\ \bibinfo {author} {\bibfnamefont {J.}~\bibnamefont
  {Chen}},\ }\bibfield  {title} {\bibinfo {title} {{Cons-training tensor
  networks: Embedding and optimization over discrete linear constraints}},\
  }\href {https://doi.org/10.21468/SciPostPhys.18.6.192} {\bibfield  {journal}
  {\bibinfo  {journal} {SciPost Phys.}\ }\textbf {\bibinfo {volume} {18}},\
  \bibinfo {pages} {192} (\bibinfo {year} {2025})}\BibitemShut {NoStop}%
\bibitem [{\citenamefont {Nakada}\ \emph {et~al.}(2025)\citenamefont {Nakada},
  \citenamefont {Tanahashi},\ and\ \citenamefont {Tanaka}}]{NaTaTa25}%
  \BibitemOpen
  \bibfield  {author} {\bibinfo {author} {\bibfnamefont {H.}~\bibnamefont
  {Nakada}}, \bibinfo {author} {\bibfnamefont {K.}~\bibnamefont {Tanahashi}},\
  and\ \bibinfo {author} {\bibfnamefont {S.}~\bibnamefont {Tanaka}},\
  }\bibfield  {title} {\bibinfo {title} {Quick design of feasible tensor
  networks for constrained combinatorial optimization},\ }\href
  {https://doi.org/10.22331/q-2025-07-21-1799} {\bibfield  {journal} {\bibinfo
  {journal} {{Quantum}}\ }\textbf {\bibinfo {volume} {9}},\ \bibinfo {pages}
  {1799} (\bibinfo {year} {2025})}\BibitemShut {NoStop}%
\bibitem [{\citenamefont {Morais}\ \emph {et~al.}(2025)\citenamefont {Morais},
  \citenamefont {Osaba}, \citenamefont {Pastor},\ and\ \citenamefont
  {Oregi}}]{MoEtAl25}%
  \BibitemOpen
  \bibfield  {author} {\bibinfo {author} {\bibfnamefont {A.}~\bibnamefont
  {Morais}}, \bibinfo {author} {\bibfnamefont {E.}~\bibnamefont {Osaba}},
  \bibinfo {author} {\bibfnamefont {I.}~\bibnamefont {Pastor}},\ and\ \bibinfo
  {author} {\bibfnamefont {I.}~\bibnamefont {Oregi}},\ }\bibfield  {title}
  {\bibinfo {title} {{Comparative Analysis of Classical and Quantum-Inspired
  Solvers: A Preliminary Study on the Weighted Max-Cut Problem}},\ }in\ \href
  {https://doi.org/10.1145/3712255.3734363} {\emph {\bibinfo {booktitle}
  {Proceedings of the Genetic and Evolutionary Computation Conference
  Companion}}},\ \bibinfo {series and number} {GECCO '25 Companion}\ (\bibinfo
  {publisher} {Association for Computing Machinery},\ \bibinfo {address} {New
  York, NY, USA},\ \bibinfo {year} {2025})\ pp.\ \bibinfo {pages}
  {2449--2457}\BibitemShut {NoStop}%
\bibitem [{\citenamefont {Rattacaso}\ \emph {et~al.}(2026)\citenamefont
  {Rattacaso}, \citenamefont {Jaschke}, \citenamefont {Ballarin}, \citenamefont
  {Siloi},\ and\ \citenamefont {Montangero}}]{RaEtAl26}%
  \BibitemOpen
  \bibfield  {author} {\bibinfo {author} {\bibfnamefont {D.}~\bibnamefont
  {Rattacaso}}, \bibinfo {author} {\bibfnamefont {D.}~\bibnamefont {Jaschke}},
  \bibinfo {author} {\bibfnamefont {M.}~\bibnamefont {Ballarin}}, \bibinfo
  {author} {\bibfnamefont {I.}~\bibnamefont {Siloi}},\ and\ \bibinfo {author}
  {\bibfnamefont {S.}~\bibnamefont {Montangero}},\ }\bibfield  {title}
  {\bibinfo {title} {Quantum algorithms for equational reasoning},\ }\href
  {https://doi.org/10.1126/sciadv.aec2736} {\bibfield  {journal} {\bibinfo
  {journal} {Sci. Adv.}\ }\textbf {\bibinfo {volume} {12}},\ \bibinfo {pages}
  {eaec2736} (\bibinfo {year} {2026})}\BibitemShut {NoStop}%
\bibitem [{\citenamefont {Garc\'{\i}a-S\'{a}ez}\ and\ \citenamefont
  {Latorre}(2012)}]{GaLa12}%
  \BibitemOpen
  \bibfield  {author} {\bibinfo {author} {\bibfnamefont {A.}~\bibnamefont
  {Garc\'{\i}a-S\'{a}ez}}\ and\ \bibinfo {author} {\bibfnamefont {J.~I.}\
  \bibnamefont {Latorre}},\ }\bibfield  {title} {\bibinfo {title} {{An exact
  tensor network for the 3SAT problem}},\ }\href
  {https://doi.org/10.26421/QIC12.3-4-8} {\bibfield  {journal} {\bibinfo
  {journal} {Quantum Inf. Comput.}\ }\textbf {\bibinfo {volume} {12}},\
  \bibinfo {pages} {283} (\bibinfo {year} {2012})}\BibitemShut {NoStop}%
\bibitem [{\citenamefont {Biamonte}\ \emph {et~al.}(2015)\citenamefont
  {Biamonte}, \citenamefont {Morton},\ and\ \citenamefont {Turner}}]{BiMoTu15}%
  \BibitemOpen
  \bibfield  {author} {\bibinfo {author} {\bibfnamefont {J.~D.}\ \bibnamefont
  {Biamonte}}, \bibinfo {author} {\bibfnamefont {J.}~\bibnamefont {Morton}},\
  and\ \bibinfo {author} {\bibfnamefont {J.}~\bibnamefont {Turner}},\
  }\bibfield  {title} {\bibinfo {title} {{Tensor Network Contractions for
  {\#}SAT}},\ }\href {https://doi.org/10.1007/s10955-015-1276-z} {\bibfield
  {journal} {\bibinfo  {journal} {J. Stat. Phys.}\ }\textbf {\bibinfo {volume}
  {160}},\ \bibinfo {pages} {1389} (\bibinfo {year} {2015})}\BibitemShut
  {NoStop}%
\bibitem [{\citenamefont {Zhu}\ and\ \citenamefont
  {Katzgraber}(2019)}]{ZhKa19}%
  \BibitemOpen
  \bibfield  {author} {\bibinfo {author} {\bibfnamefont {Z.}~\bibnamefont
  {Zhu}}\ and\ \bibinfo {author} {\bibfnamefont {H.~G.}\ \bibnamefont
  {Katzgraber}},\ }\href@noop {} {\bibinfo {title} {Do tensor renormalization
  group methods work for frustrated spin systems?}} (\bibinfo {year} {2019}),\
  \Eprint {https://arxiv.org/abs/1903.07721} {arXiv:1903.07721
  [cond-mat.dis-nn]} \BibitemShut {NoStop}%
\bibitem [{\citenamefont {Kourtis}\ \emph {et~al.}(2019)\citenamefont
  {Kourtis}, \citenamefont {Chamon}, \citenamefont {Mucciolo},\ and\
  \citenamefont {Ruckenstein}}]{KoEtAl19}%
  \BibitemOpen
  \bibfield  {author} {\bibinfo {author} {\bibfnamefont {S.}~\bibnamefont
  {Kourtis}}, \bibinfo {author} {\bibfnamefont {C.}~\bibnamefont {Chamon}},
  \bibinfo {author} {\bibfnamefont {E.~R.}\ \bibnamefont {Mucciolo}},\ and\
  \bibinfo {author} {\bibfnamefont {A.~E.}\ \bibnamefont {Ruckenstein}},\
  }\bibfield  {title} {\bibinfo {title} {Fast counting with tensor networks},\
  }\href {https://doi.org/10.21468/SciPostPhys.7.5.060} {\bibfield  {journal}
  {\bibinfo  {journal} {SciPost Phys.}\ }\textbf {\bibinfo {volume} {7}},\
  \bibinfo {pages} {060} (\bibinfo {year} {2019})}\BibitemShut {NoStop}%
\bibitem [{\citenamefont {Liu}\ \emph {et~al.}(2021)\citenamefont {Liu},
  \citenamefont {Wang},\ and\ \citenamefont {Zhang}}]{LiWaZh21}%
  \BibitemOpen
  \bibfield  {author} {\bibinfo {author} {\bibfnamefont {J.-G.}\ \bibnamefont
  {Liu}}, \bibinfo {author} {\bibfnamefont {L.}~\bibnamefont {Wang}},\ and\
  \bibinfo {author} {\bibfnamefont {P.}~\bibnamefont {Zhang}},\ }\bibfield
  {title} {\bibinfo {title} {{Tropical Tensor Network for Ground States of Spin
  Glasses}},\ }\href {https://doi.org/10.1103/PhysRevLett.126.090506}
  {\bibfield  {journal} {\bibinfo  {journal} {Phys. Rev. Lett.}\ }\textbf
  {\bibinfo {volume} {126}},\ \bibinfo {pages} {090506} (\bibinfo {year}
  {2021})}\BibitemShut {NoStop}%
\bibitem [{\citenamefont {Rams}\ \emph {et~al.}(2021)\citenamefont {Rams},
  \citenamefont {Mohseni}, \citenamefont {Eppens}, \citenamefont
  {Ja\l{}owiecki},\ and\ \citenamefont {Gardas}}]{RaEtAl21}%
  \BibitemOpen
  \bibfield  {author} {\bibinfo {author} {\bibfnamefont {M.~M.}\ \bibnamefont
  {Rams}}, \bibinfo {author} {\bibfnamefont {M.}~\bibnamefont {Mohseni}},
  \bibinfo {author} {\bibfnamefont {D.}~\bibnamefont {Eppens}}, \bibinfo
  {author} {\bibfnamefont {K.}~\bibnamefont {Ja\l{}owiecki}},\ and\ \bibinfo
  {author} {\bibfnamefont {B.}~\bibnamefont {Gardas}},\ }\bibfield  {title}
  {\bibinfo {title} {Approximate optimization, sampling, and spin-glass droplet
  discovery with tensor networks},\ }\href
  {https://doi.org/10.1103/PhysRevE.104.025308} {\bibfield  {journal} {\bibinfo
   {journal} {Phys. Rev. E}\ }\textbf {\bibinfo {volume} {104}},\ \bibinfo
  {pages} {025308} (\bibinfo {year} {2021})}\BibitemShut {NoStop}%
\bibitem [{\citenamefont {Hao}\ \emph {et~al.}(2022)\citenamefont {Hao},
  \citenamefont {Huang}, \citenamefont {Jia},\ and\ \citenamefont
  {Peng}}]{HaEtAl22}%
  \BibitemOpen
  \bibfield  {author} {\bibinfo {author} {\bibfnamefont {T.}~\bibnamefont
  {Hao}}, \bibinfo {author} {\bibfnamefont {X.}~\bibnamefont {Huang}}, \bibinfo
  {author} {\bibfnamefont {C.}~\bibnamefont {Jia}},\ and\ \bibinfo {author}
  {\bibfnamefont {C.}~\bibnamefont {Peng}},\ }\bibfield  {title} {\bibinfo
  {title} {{A Quantum-Inspired Tensor Network Algorithm for Constrained
  Combinatorial Optimization Problems}},\ }\href
  {https://doi.org/10.3389/fphy.2022.906590} {\bibfield  {journal} {\bibinfo
  {journal} {Front. Phys.}\ }\textbf {\bibinfo {volume} {10}},\ \bibinfo
  {pages} {1} (\bibinfo {year} {2022})}\BibitemShut {NoStop}%
\bibitem [{\citenamefont {Liu}\ \emph {et~al.}(2023)\citenamefont {Liu},
  \citenamefont {Gao}, \citenamefont {Cain}, \citenamefont {Lukin},\ and\
  \citenamefont {Wang}}]{LiEtAl23}%
  \BibitemOpen
  \bibfield  {author} {\bibinfo {author} {\bibfnamefont {J.-G.}\ \bibnamefont
  {Liu}}, \bibinfo {author} {\bibfnamefont {X.}~\bibnamefont {Gao}}, \bibinfo
  {author} {\bibfnamefont {M.}~\bibnamefont {Cain}}, \bibinfo {author}
  {\bibfnamefont {M.~D.}\ \bibnamefont {Lukin}},\ and\ \bibinfo {author}
  {\bibfnamefont {S.-T.}\ \bibnamefont {Wang}},\ }\bibfield  {title} {\bibinfo
  {title} {{Computing Solution Space Properties of Combinatorial Optimization
  Problems Via Generic Tensor Networks}},\ }\href
  {https://doi.org/10.1137/22M1501787} {\bibfield  {journal} {\bibinfo
  {journal} {SIAM J. Sci. Comput.}\ }\textbf {\bibinfo {volume} {45}},\
  \bibinfo {pages} {A1239} (\bibinfo {year} {2023})}\BibitemShut {NoStop}%
\bibitem [{\citenamefont {Pancotti}\ and\ \citenamefont {Gray}(2023)}]{PaGr23}%
  \BibitemOpen
  \bibfield  {author} {\bibinfo {author} {\bibfnamefont {N.}~\bibnamefont
  {Pancotti}}\ and\ \bibinfo {author} {\bibfnamefont {J.}~\bibnamefont
  {Gray}},\ }\href@noop {} {\bibinfo {title} {One-step replica symmetry
  breaking in the language of tensor networks}} (\bibinfo {year} {2023}),\
  \Eprint {https://arxiv.org/abs/2306.15004} {arXiv:2306.15004 [quant-ph]}
  \BibitemShut {NoStop}%
\bibitem [{\citenamefont {Yasuda}\ \emph {et~al.}(2024)\citenamefont {Yasuda},
  \citenamefont {Sotobayashi},\ and\ \citenamefont {Minato}}]{YaSoMi24}%
  \BibitemOpen
  \bibfield  {author} {\bibinfo {author} {\bibfnamefont {S.}~\bibnamefont
  {Yasuda}}, \bibinfo {author} {\bibfnamefont {S.}~\bibnamefont
  {Sotobayashi}},\ and\ \bibinfo {author} {\bibfnamefont {Y.}~\bibnamefont
  {Minato}},\ }\href@noop {} {\bibinfo {title} {{HOBOTAN: Efficient Higher
  Order Binary Optimization Solver with Tensor Networks and PyTorch}}}
  (\bibinfo {year} {2024}),\ \Eprint {https://arxiv.org/abs/2407.19987}
  {arXiv:2407.19987 [cs.MS]} \BibitemShut {NoStop}%
\bibitem [{\citenamefont {Gangat}\ and\ \citenamefont {Gray}(2024)}]{GaGr24}%
  \BibitemOpen
  \bibfield  {author} {\bibinfo {author} {\bibfnamefont {A.~A.}\ \bibnamefont
  {Gangat}}\ and\ \bibinfo {author} {\bibfnamefont {J.}~\bibnamefont {Gray}},\
  }\bibfield  {title} {\bibinfo {title} {Hyperoptimized approximate contraction
  of tensor networks for rugged-energy-landscape spin glasses on periodic
  square and cubic lattices},\ }\href
  {https://doi.org/10.1103/PhysRevE.110.065306} {\bibfield  {journal} {\bibinfo
   {journal} {Phys. Rev. E}\ }\textbf {\bibinfo {volume} {110}},\ \bibinfo
  {pages} {065306} (\bibinfo {year} {2024})}\BibitemShut {NoStop}%
\bibitem [{\citenamefont {Patra}\ \emph {et~al.}(2025)\citenamefont {Patra},
  \citenamefont {Singh},\ and\ \citenamefont {Or\'us}}]{PaSiOr25}%
  \BibitemOpen
  \bibfield  {author} {\bibinfo {author} {\bibfnamefont {S.}~\bibnamefont
  {Patra}}, \bibinfo {author} {\bibfnamefont {S.}~\bibnamefont {Singh}},\ and\
  \bibinfo {author} {\bibfnamefont {R.}~\bibnamefont {Or\'us}},\ }\bibfield
  {title} {\bibinfo {title} {Projected entangled pair states with flexible
  geometry},\ }\href {https://doi.org/10.1103/PhysRevResearch.7.L012002}
  {\bibfield  {journal} {\bibinfo  {journal} {Phys. Rev. Res.}\ }\textbf
  {\bibinfo {volume} {7}},\ \bibinfo {pages} {L012002} (\bibinfo {year}
  {2025})}\BibitemShut {NoStop}%
\bibitem [{\citenamefont {Ali}(2025)}]{Ma25}%
  \BibitemOpen
  \bibfield  {author} {\bibinfo {author} {\bibfnamefont {A.~M.}\ \bibnamefont
  {Ali}},\ }\href@noop {} {\bibinfo {title} {{Explicit Solution Equation for
  Every Combinatorial Problem via Tensor Networks: MeLoCoToN}}} (\bibinfo
  {year} {2025}),\ \Eprint {https://arxiv.org/abs/2502.05981} {arXiv:2502.05981
  [cs.ET]} \BibitemShut {NoStop}%
\bibitem [{\citenamefont {Dziubyna}\ \emph {et~al.}(2025)\citenamefont
  {Dziubyna}, \citenamefont {\'{S}mierzchalski}, \citenamefont {Gardas},
  \citenamefont {Rams},\ and\ \citenamefont {Mohseni}}]{DzEtAl25}%
  \BibitemOpen
  \bibfield  {author} {\bibinfo {author} {\bibfnamefont {A.~M.}\ \bibnamefont
  {Dziubyna}}, \bibinfo {author} {\bibfnamefont {T.}~\bibnamefont
  {\'{S}mierzchalski}}, \bibinfo {author} {\bibfnamefont {B.}~\bibnamefont
  {Gardas}}, \bibinfo {author} {\bibfnamefont {M.~M.}\ \bibnamefont {Rams}},\
  and\ \bibinfo {author} {\bibfnamefont {M.}~\bibnamefont {Mohseni}},\
  }\bibfield  {title} {\bibinfo {title} {{Limitations of tensor-network
  approaches for optimization and sampling: A comparison to quantum and
  classical Ising machines}},\ }\href
  {https://doi.org/10.1103/PhysRevApplied.23.054049} {\bibfield  {journal}
  {\bibinfo  {journal} {Phys. Rev. Appl.}\ }\textbf {\bibinfo {volume} {23}},\
  \bibinfo {pages} {054049} (\bibinfo {year} {2025})}\BibitemShut {NoStop}%
\bibitem [{\citenamefont {Tesoro}\ \emph {et~al.}(2026)\citenamefont {Tesoro},
  \citenamefont {Siloi}, \citenamefont {Jaschke}, \citenamefont {Magnifico},\
  and\ \citenamefont {Montangero}}]{TeEtAl26}%
  \BibitemOpen
  \bibfield  {author} {\bibinfo {author} {\bibfnamefont {M.}~\bibnamefont
  {Tesoro}}, \bibinfo {author} {\bibfnamefont {I.}~\bibnamefont {Siloi}},
  \bibinfo {author} {\bibfnamefont {D.}~\bibnamefont {Jaschke}}, \bibinfo
  {author} {\bibfnamefont {G.}~\bibnamefont {Magnifico}},\ and\ \bibinfo
  {author} {\bibfnamefont {S.}~\bibnamefont {Montangero}},\ }\bibfield  {title}
  {\bibinfo {title} {{Integer factorization via tensor-network Schnorr's
  sieving}},\ }\href {https://doi.org/10.1103/d9dl-ctt4} {\bibfield  {journal}
  {\bibinfo  {journal} {Phys. Rev. A}\ }\textbf {\bibinfo {volume} {113}},\
  \bibinfo {pages} {032418} (\bibinfo {year} {2026})}\BibitemShut {NoStop}%
\bibitem [{\citenamefont {Fishman}\ \emph
  {et~al.}(2022{\natexlab{a}})\citenamefont {Fishman}, \citenamefont {White},\
  and\ \citenamefont {Stoudenmire}}]{FiWhSt22a}%
  \BibitemOpen
  \bibfield  {author} {\bibinfo {author} {\bibfnamefont {M.}~\bibnamefont
  {Fishman}}, \bibinfo {author} {\bibfnamefont {S.}~\bibnamefont {White}},\
  and\ \bibinfo {author} {\bibfnamefont {E.~M.}\ \bibnamefont {Stoudenmire}},\
  }\bibfield  {title} {\bibinfo {title} {{The ITensor Software Library for
  Tensor Network Calculations}},\ }\href
  {https://scipost.org/10.21468/SciPostPhysCodeb.4} {\bibfield  {journal}
  {\bibinfo  {journal} {SciPost Phys. Codebases 4}\ } (\bibinfo {year}
  {2022}{\natexlab{a}})}\BibitemShut {NoStop}%
\bibitem [{\citenamefont {Fishman}\ \emph
  {et~al.}(2022{\natexlab{b}})\citenamefont {Fishman}, \citenamefont {White},\
  and\ \citenamefont {Stoudenmire}}]{FiWhSt22b}%
  \BibitemOpen
  \bibfield  {author} {\bibinfo {author} {\bibfnamefont {M.}~\bibnamefont
  {Fishman}}, \bibinfo {author} {\bibfnamefont {S.}~\bibnamefont {White}},\
  and\ \bibinfo {author} {\bibfnamefont {E.~M.}\ \bibnamefont {Stoudenmire}},\
  }\bibfield  {title} {\bibinfo {title} {{Codebase release 0.3 for ITensor}},\
  }\href {https://scipost.org/10.21468/SciPostPhysCodeb.4-r0.3} {\bibfield
  {journal} {\bibinfo  {journal} {SciPost Phys. Codebases 4-r0.3}\ } (\bibinfo
  {year} {2022}{\natexlab{b}})}\BibitemShut {NoStop}%
\bibitem [{\citenamefont {Fr\"owis}\ \emph {et~al.}(2010)\citenamefont
  {Fr\"owis}, \citenamefont {Nebendahl},\ and\ \citenamefont
  {D\"ur}}]{FrNeDu10}%
  \BibitemOpen
  \bibfield  {author} {\bibinfo {author} {\bibfnamefont {F.}~\bibnamefont
  {Fr\"owis}}, \bibinfo {author} {\bibfnamefont {V.}~\bibnamefont
  {Nebendahl}},\ and\ \bibinfo {author} {\bibfnamefont {W.}~\bibnamefont
  {D\"ur}},\ }\bibfield  {title} {\bibinfo {title} {{Tensor operators:
  Constructions and applications for long-range interaction systems}},\ }\href
  {https://doi.org/10.1103/PhysRevA.81.062337} {\bibfield  {journal} {\bibinfo
  {journal} {Phys. Rev. A}\ }\textbf {\bibinfo {volume} {81}},\ \bibinfo
  {pages} {062337} (\bibinfo {year} {2010})}\BibitemShut {NoStop}%
\bibitem [{\citenamefont {Barahona}\ \emph {et~al.}(1988)\citenamefont
  {Barahona}, \citenamefont {Gr\"{o}tschel}, \citenamefont {J\"{u}nger},\ and\
  \citenamefont {Reinelt}}]{BaEtAl88}%
  \BibitemOpen
  \bibfield  {author} {\bibinfo {author} {\bibfnamefont {F.}~\bibnamefont
  {Barahona}}, \bibinfo {author} {\bibfnamefont {M.}~\bibnamefont
  {Gr\"{o}tschel}}, \bibinfo {author} {\bibfnamefont {M.}~\bibnamefont
  {J\"{u}nger}},\ and\ \bibinfo {author} {\bibfnamefont {G.}~\bibnamefont
  {Reinelt}},\ }\bibfield  {title} {\bibinfo {title} {{An Application of
  Combinatorial Optimization to Statistical Physics and Circuit Layout
  Design}},\ }\href {https://doi.org/10.1287/opre.36.3.493} {\bibfield
  {journal} {\bibinfo  {journal} {Oper. Res.}\ }\textbf {\bibinfo {volume}
  {36}},\ \bibinfo {pages} {493} (\bibinfo {year} {1988})}\BibitemShut
  {NoStop}%
\bibitem [{\citenamefont {Lucas}(2014)}]{Lu14}%
  \BibitemOpen
  \bibfield  {author} {\bibinfo {author} {\bibfnamefont {A.}~\bibnamefont
  {Lucas}},\ }\bibfield  {title} {\bibinfo {title} {{Ising formulations of many
  NP problems}},\ }\href {https://doi.org/10.3389/fphy.2014.00005} {\bibfield
  {journal} {\bibinfo  {journal} {Front. Phys.}\ }\textbf {\bibinfo {volume}
  {2}},\ \bibinfo {pages} {5} (\bibinfo {year} {2014})}\BibitemShut {NoStop}%
\bibitem [{\citenamefont {Chertkov}\ \emph {et~al.}(2022)\citenamefont
  {Chertkov}, \citenamefont {Ryzhakov}, \citenamefont {Novikov},\ and\
  \citenamefont {Oseledets}}]{ChEtAl22}%
  \BibitemOpen
  \bibfield  {author} {\bibinfo {author} {\bibfnamefont {A.}~\bibnamefont
  {Chertkov}}, \bibinfo {author} {\bibfnamefont {G.}~\bibnamefont {Ryzhakov}},
  \bibinfo {author} {\bibfnamefont {G.}~\bibnamefont {Novikov}},\ and\ \bibinfo
  {author} {\bibfnamefont {I.}~\bibnamefont {Oseledets}},\ }\href@noop {}
  {\bibinfo {title} {{Optimization of Functions Given in the Tensor Train
  Format}}} (\bibinfo {year} {2022}),\ \Eprint
  {https://arxiv.org/abs/2209.14808} {arXiv:2209.14808 [math.NA]} \BibitemShut
  {NoStop}%
\bibitem [{\citenamefont {Press}\ \emph {et~al.}(2007)\citenamefont {Press},
  \citenamefont {Teukolsky}, \citenamefont {Vetterling},\ and\ \citenamefont
  {Flannery}}]{PrEtAl07}%
  \BibitemOpen
  \bibfield  {author} {\bibinfo {author} {\bibfnamefont {W.~H.}\ \bibnamefont
  {Press}}, \bibinfo {author} {\bibfnamefont {S.~A.}\ \bibnamefont
  {Teukolsky}}, \bibinfo {author} {\bibfnamefont {W.~T.}\ \bibnamefont
  {Vetterling}},\ and\ \bibinfo {author} {\bibfnamefont {B.~P.}\ \bibnamefont
  {Flannery}},\ }\href
  {https://www.cambridge.org/gb/universitypress/subjects/mathematics/numerical-recipes/numerical-recipes-art-scientific-computing-3rd-edition?format=HB&isbn=9780521880688}
  {\emph {\bibinfo {title} {{Numerical Recipes 3rd Edition}}}}\ (\bibinfo
  {publisher} {Cambridge University Press},\ \bibinfo {address} {Cambridge,
  UK},\ \bibinfo {year} {2007})\BibitemShut {NoStop}%
\end{thebibliography}%

\end{document}